\documentclass[prl,onecolumn,superscriptaddress,nofootinbib]{revtex4}

\usepackage{graphicx}
\usepackage{natbib}
\usepackage{color}
\usepackage{enumerate}
\usepackage{epsfig,subfigure}
\usepackage{amsmath,amssymb,latexsym}
\usepackage{ascmac}
\usepackage{bm}
\def\U#1{{\rm #1}} 

\newtheorem{theorem}{{\bf Theorem}}

\newtheorem{pro}{{\bf Proposition}}
\newcommand{\bra}[1]{\langle #1 |}
\newcommand{\ket}[1]{| #1 \rangle}

\newcommand{\expect}[1]{\left\langle #1 \right\rangle} 

\newcommand{\sq}{\qquad $\blacksquare$}

\begin{document}
\title
{Quantum key distribution with setting-choice-independently correlated light sources}
      \author{Akihiro Mizutani\footnote{mizutani@qi.mp.es.osaka-u.ac.jp}}
\affiliation{Graduate School of Engineering Science, Osaka University,
  Toyonaka, Osaka 560-8531, Japan}
\author{Go~Kato}
\affiliation{NTT Communication Science Laboratories, NTT Corporation, 3-1,
  Morinosato Wakamiya Atsugi-Shi, Kanagawa, 243-0198, Japan}
\affiliation{NTT Research Center for Theoretical Quantum Physics, NTT Corporation, 3-1 Morinosato-Wakamiya, Atsugi, Kanagawa 243-0198, Japan}
\author{Koji~Azuma}
\affiliation{NTT Basic Research Laboratories, NTT Corporation, 
  3-1, Morinosato-Wakamiya Atsugi-Shi, 243-0198, Japan}
\affiliation{NTT Research Center for Theoretical Quantum Physics, NTT Corporation,
  3-1 Morinosato-Wakamiya, Atsugi, Kanagawa 243-0198, Japan}
\author{Marcos Curty}
\affiliation{EI Telecomunicaci\'on, Department of Signal Theory and Communications, University of Vigo, Vigo E-36310, Spain}
      \author{Rikizo Ikuta}
\affiliation{Graduate School of Engineering Science, Osaka University,
  Toyonaka, Osaka 560-8531, Japan}
      \author{Takashi Yamamoto}
\affiliation{Graduate School of Engineering Science, Osaka University,
  Toyonaka, Osaka 560-8531, Japan}
      \author{Nobuyuki Imoto}
\affiliation{Graduate School of Engineering Science, Osaka University,
  Toyonaka, Osaka 560-8531, Japan}
\author{Hoi-Kwong Lo}
\affiliation{
Center for Quantum Information and Quantum Control, Department of Physics and Dept. of Electrical \& Computer Engineering, 
University of Toronto, M5S 3G4 Toronto, Canada}
\author{Kiyoshi~Tamaki}
\affiliation{NTT Basic Research Laboratories, NTT Corporation, 
  3-1, Morinosato-Wakamiya Atsugi-Shi, 243-0198, Japan}
\affiliation{Graduate School of Science and Engineering for Education, University of
Toyama, Gofuku 3190, Toyama, 930-8555, Japan}
\begin{abstract}
  {
Despite the enormous theoretical and experimental progress made so far in quantum key distribution (QKD), the security of most existing QKD implementations is not rigorously established yet.
A critical obstacle is that almost all existing security proofs make ideal assumptions on the QKD devices. 
Problematically, such assumptions are hard to satisfy in the experiments, and therefore it is not obvious how to apply such security proofs to practical QKD systems. 
Fortunately, any imperfections and security-loopholes in the measurement devices can be perfectly closed by measurement-device-independent QKD (MDI-QKD),
and thus we only need to consider how to secure the source devices.
Among imperfections in the source devices, correlations between the sending pulses are one of the principal problems. 
In this paper, we consider a setting-choice-independent correlation (SCIC) framework
in which the sending pulses can present arbitrary correlations but they are independent of the previous setting choices
such as the bit, the basis and the intensity settings. Within the framework of SCIC,
  we consider the dominant fluctuations of the sending states, such as the relative phases and the intensities,
  and provide a self-contained information theoretic security proof for the loss-tolerant QKD protocol
  in the finite-key regime. We demonstrate the feasibility of secure quantum communication within a reasonable number of pulses sent,
  and thus we are convinced that our work constitutes a crucial step toward guaranteeing implementation security of QKD.}
\end{abstract}

\maketitle

\section{Introduction}
\label{Introduction}
Quantum key distribution (QKD)~\cite{LoNphoto2014} is one of the most promising applications of quantum information processing, 
and it is now on the verge of global commercialization. 
Nonetheless, there are still several theoretical and experimental challenges~\cite{Diamanti16} that need to be addressed before its wide-scale deployment.
One such challenge is the lack of practical security proofs that bridge the gap between theory and practice.
In the security proof of QKD, one typically assumes some mathematical models for Alice and Bob's devices.
However, if these models do not faithfully capture the physical properties of the actual QKD devices, the security of the systems is no longer guaranteed.
In fact, such discrepancies between device models assumed in security proofs and the properties of actual devices 
could be exploited by Eve to attack both the source~\cite{PhysRevA.91.032326,PhysRevA.92.022304} and the detection
apparatuses~\cite{2006PhysRevA.74.022313,Bing07,Lamas-Linares:07,2008PhysRevA.78.019905,2008PhysRevA.78.042333,Lydersen2010,Gerhardt2011,20111367-2630-13-7-073024}.
It is therefore indispensable for realising secure QKD to develop security proof techniques that can
be applied to actual devices. 

One possible approach to close this gap is to use device-independent QKD
~\cite{yao1998,PhysRevLett.98.230501,PhysRevLett.113.140501,Rotem16}.
Its main drawback is, however, that it delivers a quite low secret key rate with current technology,
and it still requires some device characterisations
\footnote{
  Note that device-independent QKD is known to be vulnerable to memory attacks~\cite{PhysRevLett.110.010503}. 
  See also~\cite{curty2017} for a possible countermeasure against this type of attacks. }. 
An alternative solution is to use measurement-device-independent (MDI) QKD~\cite{PhysRevLett.108.130503}, which guarantees 
the security of QKD without making any assumption on the measurement device.
That is, MDI-QKD completely closes the security loophole in the detection unit.
This technique still requires, however, that certain assumptions on the source device are satisfied.

Unfortunately, the status of the security proofs with practical light sources is not fully satisfactory, as so far
only a few security proofs accommodate realistic imperfections in the source device. 
  Among the imperfections in the source, one of the crucial problems is the presence of correlations among the sending pulses.
  We categorize these correlations into two types: the first type is the {\it setting-choice-independent correlation} (SCIC)
  where the correlation is independent of Alice's choices of settings such as the bit, the basis, and the intensity settings,
  and the second type is the {\it setting-choice-dependent correlation} (SCDC) where the correlation is dependent on Alice's
  setting choices.
  For instance, the former case (SCIC) may arise when the temperature in the laser drifts slowly over time due to thermal effects,
  where such drift could depend on how long we have operated a device and the ambient temperature of the room. 
  Another example may be found in modulation devices which are operated by power supply fluctuating in time.
  On the other hand, the latter case (SCDC) occurs when
  the $i^{\U{th}}$ sending state could depend on the previous setting choices that Alice
  has made up to the $(i-1)^{\U{th}}$ pulse.
  That is to say, secret information encoded in the previous quantum signals sent by Alice could be leaked to subsequent quantum signals
  sent by Alice. In other words, subsequent signals could act as side channels for previous signals. 
  Recently, the SCDC between the intensities of different pulses 
  has been observed experimentally~\cite{qcrypt2017tomita}.
  Also, the authors of~\cite{qcrypt2017tomita} conducted a security analysis which is valid for the restricted scenario where
  only the nearest neighbour correlation is considered. 
  More in general, however, the $i^{\U{th}}$ state could be dependent on all the previous setting choices that Alice has made. 
  This general correlation seems to be very hard to deal with theoretically, and even if we would have
  a theoretical countermeasure against it, the characterisation of the device might be highly non-trivial.
  Fortunately, it would be reasonable to assume that the SCDC could be eliminated if the modulation devices
  are initialized each time after Alice emits a pulse.
  For instance, before Alice sends the $(i+1)^{\U{th}}$ pulse, she applies a
  random voltage to the modulation devices several times until the
  setting-choice information up to the $i^{\U{th}}$ pulse which is stored in the device is deleted.
  This potential solution may decrease the repetition rate of the source, however, this could be overcome by multiplexing several
  sources,
  for instance by employing integrated silicon photonics~\cite{Ding2017,Sibson:17,Ma:16}.
  With this reasonable solution, we are left with rigorously dealing with the SCIC.

    In this paper, we consider the dominant fluctuations of the sending state, such as the relative
    phase~\cite{AkihiroNJP2015,PhysRevA.93.042325,PhysRevA.92.032305,PhysRevA.93.042308}
    and the intensity~\cite{PhysRevA.77.042311,AkihiroNJP2015,Hayashi2016Decoy,qcrypt2017tomita}
    within the framework of SCIC, and we provide an information theoretic security proof in the finite-key regime. 
    In particular, we consider the loss-tolerant QKD protocol~\cite{PhysRevA.90.052314} that is a
    BB84 type protocol which, unlike the standard BB84 protocol~\cite{bb84}, has
    the advantage of being robust against phase modulation errors. 
    The loss-tolerant protocol is highly practical and has been experimentally demonstrated in both
    prepare \& measure QKD~\cite{PhysRevA.92.032305,Zbinden18} and MDI-QKD~\cite{PhysRevA.93.042308}.
Our main contribution is to explicitly write down all the assumptions that we impose on QKD systems, and by using only these assumptions we give a self-contained 
security proof. 
  Our numerical simulations of the key generation rate show that provably secure keys can be distributed over long distance within a reasonable number of pulses sent, e.g, $10^{12}$ pulses.

The paper is organised as follows. In section~\ref{Assumptions and protocol description}, we describe the assumptions that we make on Alice and Bob's devices
and we introduce the protocol considered. 
In section~\ref{Secret key generation length}, we present a formula for the key generation length of the protocol.
This formula depends on certain parameters that need to be estimated; the estimation results for these parameters are shown in section~\ref{Parameter estimations}.
Then in section~\ref{keyrate}, we present our numerical simulation results for the key generation rate. Here, we assume realistic intervals for the actual phases and intensities under
the framework of SCIC,
and we show that secure communication is possible within a reasonable time frame of signal transmission, say $10^{12}$ signals.
Finally, section~\ref{Conclusion} summarises the paper. There are also various appendixes describing all the detailed derivations of the parameters in the key rate.

        \section{Assumptions and protocol description}
        \label{Assumptions and protocol description}
        Here, we introduce the assumptions on Alice and Bob's devices and the protocol we consider throughout this paper.
        To describe the assumptions, we use a shorthand notation
        $\bm{X}^{i}:=X^{i},X^{i-1},...,X^{1}$ for a sequence of random variables $\{X^j\}^i_{j=1}$ and $X^0:=0$.
        In what follows, we first summarise the assumptions we make on the sending devices as well as those on the measurement devices,
        and then we move on to the description of the protocol.

        \begin{figure}[t]
\includegraphics[width=8.2cm]{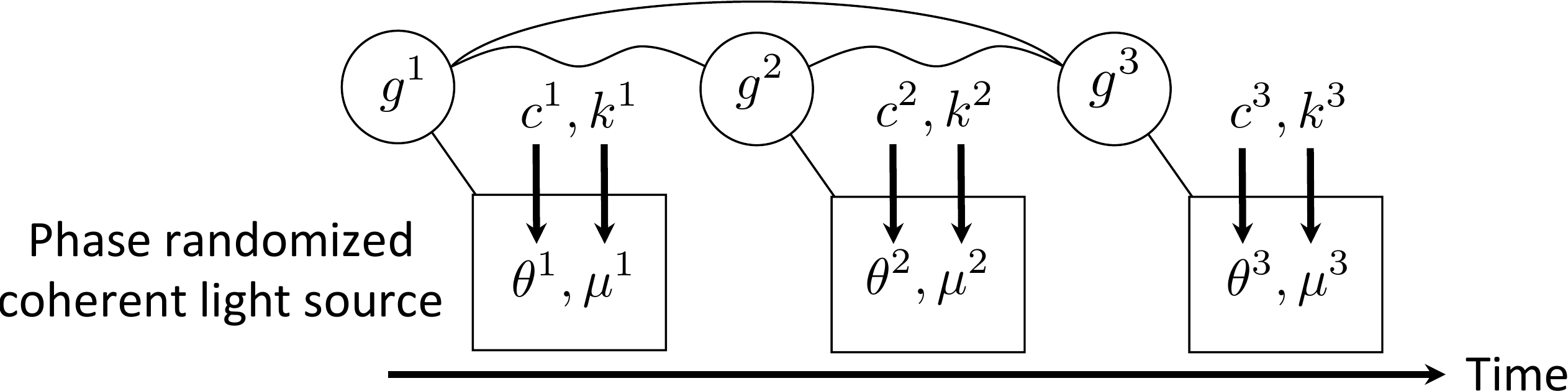}
\caption{
A phase randomized coherent light source with SCIC (with $N_{\U{sent}}=3$). 
  The internal states of the source device $\{g^i\}^{N_{\U{sent}}}_{i=1}$ that determine $\{\theta^i\}^{N_{\U{sent}}}_{i=1}$ and
  $\{\mu^i\}^{N_{\U{sent}}}_{i=1}$
  are setting-choice-independently correlated [see assumption (A-2)]. In each trial, Alice inputs $c^i$ and $k^i$ to the source device,
  and depending on these choices and $g^i$, the phase $\theta^i$ and the intensity $\mu^i$ are determined. 
  Importantly, the internal states of the source device $\{g^i\}^{N_{\U{sent}}}_{i=1}$ can be arbitrary correlated with each other. 
  Note that the secret information contained in previous signals (namely, $\bm{c}^{i-1}$ and $\bm{k}^{i-1}$) is {\it not} leaked to subsequent signals. This avoids
  the side channel problem.
}
\label{fig:SIC}
        \end{figure}
\begin{figure}[t]
\includegraphics[width=8.2cm]{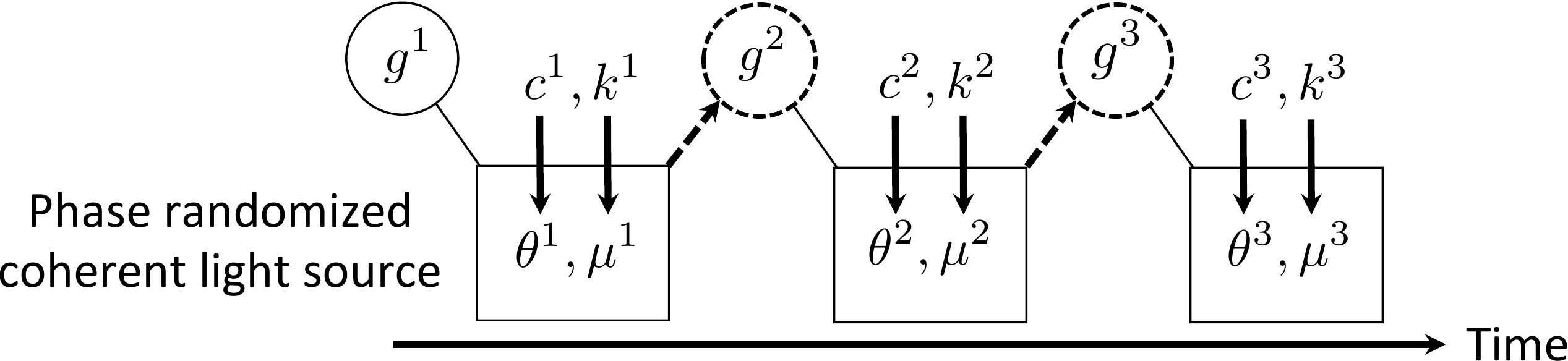}
\caption{
  This figure exemplifies the SCDC that is forbidden in our security assumptions (with $N_{\U{sent}}=3$). 
  It shows that the $i^{\U{th}}$ internal state of the source device $g^i$ depends on the previous outcomes $\bm{\theta}^{i-1}$ and
  $\bm{\mu}^{i-1}$.
  In this case, the secret information contained in the previous quantum signals (namely, $\bm{c}^{i-1}$ and $\bm{k}^{i-1}$) 
    could be leaked to the $i^{\U{th}}$ quantum signal sent by Alice.
    In other words, the $i^{\U{th}}$ sending signal could act as a side channel for the previous $(i-1)^{\U{th}}$ signals.
    }
  \label{fig:koashi}
\end{figure}
                
                \subsection{Assumptions on Alice's transmitter}
                \label{Assumptions on the devices}
                (A-1) {\it Assumption on the sending state $\hat{\rho}^{i}(\theta^i,\mu^i)_{B^i}$}

                We assume that Alice employs a coherent light source with a Poissonian photon number distribution in any basis, bit and intensity setting.
                Here we denote by $c^{i}\in\mathcal{C}:=\{0_Z,1_Z,0_X\}$ Alice's bit and basis choice for the $i^{\U{th}}$ pulse, and
                $k^{i}\in\mathcal{K}:=\{k_1,k_2,k_3\}$ Alice's intensity setting choice for the $i^{\U{th}}$ pulse
                \footnote{The method introduced in this paper is general and can be applied to many different protocols.
                  To simplify the discussion, however, we consider the loss-tolerant three-state protocol~\cite{PhysRevA.90.052314} with two decoy states.}.
                Moreover, we consider that the $i^{\U{th}}$ signals are in a single-mode
                \footnote{The single-mode condition means that each $i^{\U{th}}$ emitted signal can be mathematically
                  characterised by a single creation operator. This creation operator can, however, be different for different pulses.}.
                Also, we assume that she uses phase encoding, {\it i.e.}, she encodes the $i^{\U{th}}$
                bit and basis information into the relative phase $\theta^i$ between two pulses,
                a signal and a reference pulse, whose common phase $\delta\in[0,2\pi)$ is perfectly randomized.
                  We denote by $\mu^{i}$ the $i^{\U{th}}$ actual value of the intensity generated by Alice's source.
                  Mathematically, given the phase and the intensity, the $i^{\U{th}}$
                  sending state $\hat{\rho}^{i}(\theta^i,\mu^i)_{B^i}$ to Bob in the system $B^i$
                  can be described as
  \begin{align}
    \hat{\rho}^{i}(\theta^i,\mu^i)_{B^i}&=\frac{1}{2\pi}\int^{2\pi}_{0}\hat{P}[\ket{e^{\U{i}(\delta+\theta^{i})}
        \sqrt{\mu^{i}/2}}_{S^i}\ket{e^{\U{i}\delta}\sqrt{\mu^{i}/2}}_{R^i}]d\delta.
\label{phaserandom}
  \end{align}
Here, we assume that the intensities of the signal and the reference pulses are the same and equal to $\mu^i/2$
\footnote
    {This assumption is not mandatory. Even if the intensities of the $i^{\U{th}}$ signal and reference
      pulses differ, the security proof can be established by introducing an additional filter operation in the security proof as explained in~\cite{PhysRevA.90.052314}.    }. 
In Eq.~(\ref{phaserandom}), we define $\hat{P}[\ket{\cdot}]:=\ket{\cdot}\bra{\cdot}$, the subscripts 
$S^i$ and $R^i$ respectively represent the optical 
modes of the $i^{\U{th}}$ signal and reference pulse, and $\ket{e^{\U{i}\theta}\sqrt{\mu}}_{S^i(R^i)}$ denotes a coherent state in mode $S^i(R^i)$, {\it i.e.,}
$\ket{e^{\U{i}\theta}\sqrt{\mu}}_{S^i(R^i)}=e^{-\mu/2}\sum^{\infty}_{n=0}(e^{\U{i}\theta}\sqrt{\mu})^n\ket{n}_{S^i(R^i)}/\sqrt{n!}$ 
with $\ket{n}_{S^i(R^i)}$ being a Fock state with $n$ photons in mode $S^i(R^i)$. Eq.~(\ref{phaserandom}) can be rewritten as 
\begin{align}
  \hat{\rho}^{i}(\theta^i,\mu^i)_{B^i}=\sum^{\infty}_{n^i=0}p(n^i|\mu^i)\hat{P}[\ket{\hat{\Upsilon}^{i}(\theta^{i},n^{i})}_{B^i}],
  \label{phaserandom2}
\end{align}
where $p(n^i|\mu^i):=e^{-\mu^i}(\mu^i)^{n^i}/n^i!$ and the $n^i$-photon state $\hat{P}[\ket{\hat{\Upsilon}^{i}(\theta^{i},n^{i})}_{B^i}]$
is defined as
\begin{align}
  \hat{P}[\ket{\hat{\Upsilon}^{i}(\theta^{i},n^{i})}_{B^i}]
  :=\frac{\hat{N}^{n^i}_{B^i}\hat{\rho}^{i}(\theta^{i},\mu^{i})_{B^i}\hat{N}^{n^i}_{B^i}}
  {\U{tr}[\hat{N}^{n^i}_{B^i}\hat{\rho}^{i}(\theta^{i},\mu^{i})_{B^i}]},
      \label{Upsilon}
\end{align}
where $\hat{N}^{n^i}_{B^i}:=\sum^{n^i}_{k=0}\hat{P}[\ket{n^i-k}_{S^i}\ket{k}_{R^i}]$. 

Furthermore, we suppose that there are no side-channels in Alice's source and Eve can only manipulate Bob's system $B$ with
her arbitrary prepared ancilla.

In the following, we first explain our correlation model for the source device, and we make assumptions on how the phases $\{\theta^i\}^{N_{\U{sent}}}_{i=1}$ and the intensities
$\{\mu^i\}^{N_{\U{sent}}}_{i=1}$ are determined in the source device, where $N_{\U{sent}}$ denotes the number of pulse pairs
(signal and reference pulses) sent by Alice. 
See Fig.~\ref{fig:SIC} for a schematic explanation of our correlation model of the source device.
For illustration purposes, we exemplify in Fig.~\ref{fig:koashi} the setting-choice-dependent correlation (SCDC)
that is not taken into account in our security analysis. \\
        (A-2) {\it Assumption on the correlation: Setting-choice-independent correlation (SCIC)}

The correlation model we consider is the setting-choice-independent correlation (SCIC),
which means that the internal state of the source device which determines the $i^{\U{th}}$ sending state 
is arbitrarily correlated with the previous internal states of the source device but it does not depend on the previous
setting choices made by Alice.
We denote by $g^i$ the classical random variable representing the $i^{\U{th}}$ internal state of the source device; it 
determines the correspondence between the setting choices 
($c^i$ and $k^i$) and the output parameters from the source device ($\theta^i$ and $\mu^i$). 
We suppose that $g^i$ depends on the past internal state of the source device $\bm{g}^{i-1}$ and is independent of the past
setting choices and output parameters
\footnote{
  Note that since the output parameters ($\theta^i$ and $\mu^i$)
  have the information of the setting choices ($c^i$ and $k^i$), we also need to impose the independence of
  $g^i$ from $\bm{\theta}^{i-1}$ and $\bm{\mu}^{i-1}$. 
  }. 
Hence, if we denote the $i^{\U{th}}$ setting choices and output parameters by
        \begin{align}
  P^{i}:=(\theta^{i},\mu^{i},c^{i},k^{i}),
        \end{align}
        the SCIC model can be mathematically expressed in terms of a probability distribution satisfying 
        for any $\bm{P}^{i-1}$ and $\bm{g}^{i-1}$ the following
      \begin{align}
        p(g^{i}|\bm{g}^{i-1},\bm{P}^{i-1})=p(g^{i}|\bm{g}^{i-1}).
        \label{assumpA1}
      \end{align}
      (A-3) {\it Assumption on the random choice of $c^{i}$ and $k^{i}$}

      We assume that conditioned on the past realisation $\bm{P}^{i-1}$ and $\bm{g}^{i}$, then $c^{i}$ and $k^{i}$ are independent of each 
      other and also independent of $\bm{P}^{i-1}$  and of $\bm{g}^{i}$, which is expressed by the following condition
\begin{align}
  p(c^{i},k^{i}|\bm{g}^{i},\bm{P}^{i-1})=p(c^{i})p(k^{i}).
  \label{randomchoiceCK}
\end{align}
(A-4) {\it Assumption on the independence of $\theta^{i}$ and $\mu^{i}$}

We suppose that the phase $\theta^{i}$ (intensity $\mu^{i}$) only depends on the setting choice $c^{i}$ ($k^{i}$) and on 
$g^{i}$. Mathematically, this means that the probability distributions satisfy
\begin{align}
  p(\theta^{i},\mu^{i}|c^{i},k^{i},\bm{g}^{i},\bm{P}^{i-1})=
  p(\theta^{i}|c^{i},g^{i})p(\mu^{i}|k^{i},g^{i}).
    \label{thetamu}
\end{align}
(A-5) {\it Assumption on unique determination of $\theta^i$ and $\mu^i$}

The phase $\theta^{i}$ (intensity $\mu^{i}$) is uniquely determined given $g^i$ and the setting choice $c^i$ ($k^i$) as $\theta^i_{c^i,g^{i}}$ ($\mu^i_{k^i,g^{i}}$), that is, $\theta^{i}$ ($\mu^i$) is a function of $c^i$ ($k^i$) and $g^i$. This is expressed as
\begin{align}
  p(\theta^i|c^i,g^{i})=\delta(\theta^i,\theta^i_{c^i,g^{i}}),~~
  p(\mu^i|k^i,g^{i})=\delta(\mu^i,\mu^i_{k^i,g^{i}}),
  \label{unique}
\end{align}
where $\delta(x,y)$ denotes the Kronecker delta.
Note that Eq.~(\ref{unique}) does not impose any restriction on $\{g^i\}^{N_{\U{sent}}}_{i=1}$ since there exists the information of 
$\theta^i$ and $\mu^i$ somewhere in the source device, and we can take the parameters $\{g^i\}^{N_{\U{sent}}}_{i=1}$ such that $\{g^i\}^{N_{\U{sent}}}_{i=1}$
uniquely determine the correspondence between $\{c^i\}_{c^i\in\mathcal{C}}$ and $\{\theta^i_{c^i,g^{i}}\}_{c^i\in\mathcal{C}}$,
  and $\{k^i\}_{k^i\in\mathcal{K}}$ and $\{\mu^i_{k^i,g^{i}}\}_{k^i\in\mathcal{K}}$. 

For the security analysis, we define the random variable associated to tagged events as follows.
\\
(D-1) {\it Definition of tagged random variable}

For the internal state of the source device $g^i$, we define the {\it untagged} set $\mathcal{G}^i_{\U{unt}}$ as 
\begin{align}
  \mathcal{G}^i_{\U{unt}}=\{g^i|\forall c^i\in\mathcal{C},~\forall k^i\in\mathcal{K},~\theta^i_{c^i,g^i}\in R^{c^i}_{\U{ph}},~\mu^i_{k^i,g^i}\in R^{k^i}_{\U{int}}\},
\end{align}
and if $g^i\in (\notin)\mathcal{G}^i_{\U{unt}}$, we call the $i^{\U{th}}$ pulse the untagged (tagged) signal, which we denote by $t^i=u~(t)$. In the above definition of the untagged set,
$R^{c^i}_{\U{ph}}$ and $R^{k^i}_{\U{int}}$ respectively denote the interval of the phase for $c^i$ and the interval of the intensity for $k^i$.
\\
(A-6) {\it Assumption on the intervals for the phase and intensity}

The $i^{\U{th}}$ interval of the phase $R^{c^i}_{\U{\U{ph}}}$ is assumed to be given by
\begin{align}
R^{0_Z}_{\U{\U{ph}}}=[\theta^{\U{L}}_{0_Z},\theta^{\U{U}}_{0_Z}],~~
R^{1_Z}_{\U{\U{ph}}}=[\theta^{\U{L}}_{1_Z},\theta^{\U{U}}_{1_Z}],~~
R^{0_X}_{\U{\U{ph}}}=[\theta^{\U{L}}_{0_X},\theta^{\U{U}}_{0_X}]
  \label{Rph}
\end{align}
for all instances $i$, where $R^{0_Z}_{\U{\U{ph}}}$, $R^{1_Z}_{\U{\U{ph}}}$ and $R^{0_X}_{\U{\U{ph}}}$ do not overlap each other 
and the parameters $\{\theta^{\U{L}}_c\}_{c\in\mathcal{C}}$ and $\{\theta^{\U{U}}_c\}_{c\in\mathcal{C}}$ must satisfy 
  $-\frac{\pi}{6}<\theta^{\U{L}}_{0_Z}\le0$, $0\le\theta^{\U{U}}_{0_Z}<\frac{\pi}{6}$,
$\frac{5\pi}{6}<\theta^{\U{L}}_{1_Z}\le\pi$, $\pi\le\theta^{\U{U}}_{1_Z}<\frac{7\pi}{6}$,
$\frac{\pi}{3}<\theta^{\U{L}}_{0_X}\le\frac{\pi}{2}$, and $\frac{\pi}{2}\le\theta^{\U{U}}_{0_X}<\frac{2\pi}{3}$. 
Also, the $i^{\U{th}}$ interval of the intensity $R^{k^i}_{\U{\U{int}}}$ has the form 
\begin{align}
  R^{k}_{\U{\U{int}}}=[\mu_k^-,\mu_k^+],
  \label{Rint}
\end{align}
for all instances $i$, and we suppose that the following three conditions are satisfied
\footnote{These conditions are needed in the decoy-state method that is used for the parameter estimation
(see Appendix~\ref{sec:appC} for details). }: 
$\mu^+_{k_3}< \mu^-_{k_2}$, $\mu^+_{k_2}+\mu^+_{k_3}<\mu^-_{k_1}$ and $\mu^+_{k_1}\leq 1$.
\\
(A-7) {\it Assumption on the number of tagged signals}

We define the {\it good} set $\bm{\mathcal{G}}^{N_{\U{sent}}}_{\U{good}}$ of $\bm{g}^{N_{\U{sent}}}$ as that whose number of tagged events
$n_{\U{tag}}:=|\{i|g^i\notin\mathcal{G}^i_{\U{unt}}\}|$ is upper bounded by a constant number $N_{\U{tag}}$ as
 \begin{align}
   \bm{\mathcal{G}}^{N_{\U{sent}}}_{\U{good}}:=\{\bm{g}^{N_{\U{sent}}}|n_{\U{tag}}\le N_{\U{tag}}\}.
   \label{Ntagdef}
 \end{align}
 We suppose that the probability of $\bm{g}^{N_{\U{sent}}}$ not being an element of $\bm{\mathcal{G}}^{N_{\U{sent}}}_{\U{good}}$ is upper bounded by $p_{\U{fail}}$,
 which is expressed as
     \begin{align}
       \sum_{\bm{g}^{N_{\U{sent}}}\notin \bm{\mathcal{G}}^{N_{\U{sent}}}_{\U{good}}}p(\bm{g}^{N_{\U{sent}}})\le p_{\U{fail}}.
       \label{tagfail}
     \end{align}

     \subsection{Assumptions on Bob's measurement unit}
     \label{assbob}
(B-1) {\it Assumption on basis-independent detection efficiency}

  We denote by $\{\hat{M}_{y^{i},b^{i}}\}_{y^{i}\in\{0,1,\emptyset\}}$ the $i^{\U{th}}$ POVM (positive operator-valued measure)
  for Bob's measurement in the basis $b^{i}\in\mathcal{B}:=\{Z,X\}$,
  where $\hat{M}_{0,b^{i}}$ ($\hat{M}_{1,b^{i}}$) represents the POVM element associated to the detection of the bit value $y^{i}=0~(1)$
  in the basis $b^{i}$, and the element $\hat{M}_{\emptyset,b^{i}}$ 
  represents the failure of outputting a bit value.
  We suppose that whether a detection occurs or not for each pulse pairs does not depend on the chosen measurement basis $b^{i}$;
  this condition is represented as
          \begin{align}
\hat{M}_{\emptyset}:=\hat{M}_{{\emptyset},Z}=\hat{M}_{\emptyset,X}.
            \label{independentefficiency}
          \end{align}
     (B-2) {\it Assumption on random choice of the measurement basis}

          We assume that Bob measures each incoming signal in a basis $b^{i}\in\mathcal{B}$ chosen independently of
          the previous basis choices and measurement outcomes. This condition is expressed in terms of the probability distribution as
\begin{align}
  p(b^{i}|\bm{b}^{i-1},\bm{y}^{i-1})=p(b^{i}).
  \label{radnomb}
\end{align}
          Furthermore, we suppose that there are no side-channels in Bob's measurement device.
     Let us remark that our security model allows the use of threshold detectors; this simply implies that Bob's $Z$ and $X$
     basis measurements are not necessarily measurements on a qubit space.
     Note also that any error in the detection apparatus (say, for example, modulation errors)
     can be accommodated in our security proof as long as the assumptions stated in (B-1) and (B-2) are satisfied.

     \subsection{Protocol description}
     \label{Protocol description}
     \begin{figure}[t]
\includegraphics[width=8cm]{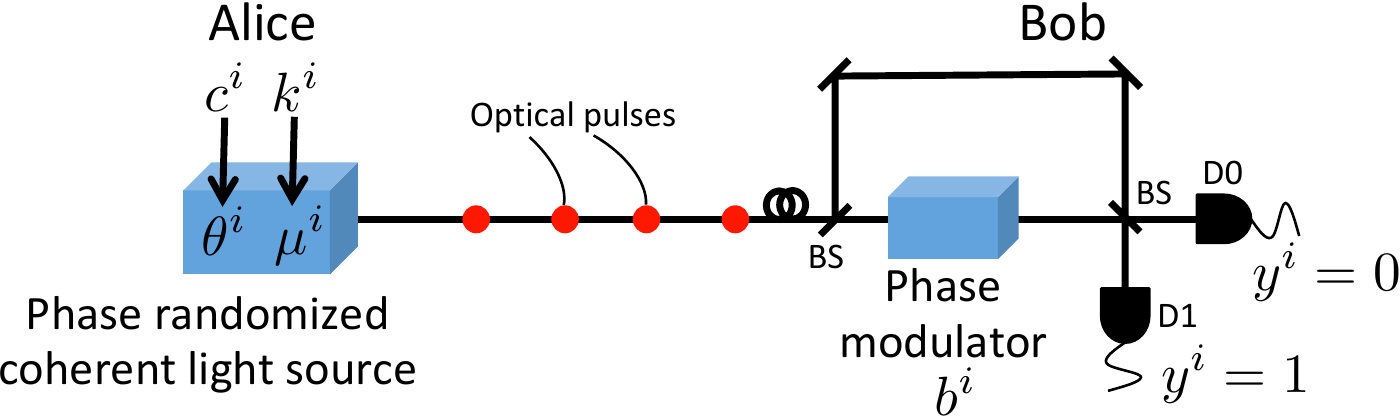}
\caption{
Description of the actual protocol with a typical measurement setup. In the $i^{\U{th}}$ trial (with $1\leq i\leq {N_{\U{sent}}}$), Alice's source device emits two consecutive coherent pulses: a signal and a reference pulse. 
  Alice first inputs the basis and bit information $c^{i}\in\mathcal{C}$ and the intensity setting $k^{i}\in\mathcal{K}$ that she selects probabilistically. 
  Let $\theta^{i}$ and $\mu^{i}$ denote the relative phase between the signal and the reference pulses and the total actual intensity of both pulses, respectively. 
  On the receiving side, Bob uses a 50:50 beamsplitter (BS) to split the received pulses into two beams. Afterward, he
  applies a phase shift 0 or $\pi/2$ to one of them according to his basis choice $b^{i}=Z$ or $b^{i}=X$, respectively.
  The pulses are then recombined at a 50:50 BS. A click in the detector D0 (D1) provides Bob the bit $y^{i}=0$ $(y^{i}=1)$. }
\label{fig:actual}
     \end{figure}
             In this section, we describe the protocol of which we prove the security.
             See Fig.~\ref{fig:actual} for a typical setup of the actual protocol. 
In particular, we consider the loss-tolerant protocol~\cite{PhysRevA.90.052314}. 
Also, we suppose that Alice uses the decoy-state method~\cite{PhysRevLett.91.057901,PhysRevLett.94.230504,PhysRevLett.94.230503}
with one signal and two decoys, 
and we consider asymmetric coding, {\it i.e.,} the $Z$ and $X$ bases are chosen with probabilities $p^{A}_Z:=\sum_{c^i=0_Z,1_Z}p(c^i)$
and $p^{A}_X:=p(c^i=0_X)$, respectively. In addition, we assume that the secret key is generated from those events
where both Alice and Bob select the $Z$ basis regardless of their intensity settings. 

Next, we show in detail how the protocol runs.
In its description, $|A|$ represents the cardinality of a set or length of a bit string
depending on whether $A$ is a set or a bit string, respectively. 
The protocol is composed of the following steps:
\\\\
    {0. \it Device characterisation and protocol parameter choice}
    \\
First, Alice characterises her source to determine the value of the parameters 
$R^{c}_{\U{\U{ph}}}$, $R^{k}_{\U{\U{int}}}$ for all $c\in\mathcal{C}$ and $k\in\mathcal{K}$, $N_{\U{tag}}$ and $p_{\U{fail}}$.
Also, Alice and Bob decide the secrecy parameter $\epsilon_{\U{s}}$ given by Eq.~(\ref{secrecy}), 
the correctness parameter $\epsilon_{\U{c}}$, the upper bound on $N_{\U{sent}}$ which we shall denote by $N$,
and the quantity $N_{\U{det}}$ that is associated to the termination condition.

After this characterisation step, Alice (Bob) repeats the following step 1 (step 2) and both Alice and Bob repeat step 3 for
$i = 1,...,N_{\U{sent}}$ until the condition in the sifting step is met
\footnote{
  Note that we adopt the {\it iterative sifting} procedure with a basis independent termination condition, which
  has been recently analysed in~\cite{KiyoQST2017}.
  In this procedure, after each quantum transmission round, Bob announces whether or not the received signal produced a detection click
  in his measurement apparatus. 
  And, in the case of a detection click, both Alice and Bob announce their basis choices, and Bob also declares his
  measurement outcome except for the event where both of them selected the $Z$ basis.
  Then, the quantum communication part of the protocol terminates when the basis independent termination condition is satisfied. }.
\\\\
    {1. \it Preparation}
    \\
    For each $i$, Alice randomly selects the intensity setting $k^{i}\in\mathcal{K}$ with 
    probabilities $p_{k_1}:=p(k^i=k_1)$, $p_{k_2}:=p(k^i=k_2)$ 
    and $p_{k_3}:=p(k^i=k_3)$, and the basis $a^{i}\in\mathcal{B}$ with probabilities $p^{A}_Z$ and
    $p^{A}_X=1-p^{A}_Z$. Afterward, if $a^{i}=Z$ she chooses the bit information with probability $1/2$;
    otherwise, she chooses $c^{i}=0_X$. Finally, she generates the signal and reference pulses according to her choice of $k^{i}$
    and $c^{i}$, and sends them to Bob via a quantum channel.
\\\\
    {2. \it Measurement}
    \\
Bob measures the incoming signal and reference pulses using the measurement basis $b^{i}\in {\mathcal B}$, 
which he selects with probabilities $p^{B}_Z:=p(b^i=Z)$ and $p^{B}_X:=p(b^i=X)$. 
The outcome is recorded as $\{0, 1, \perp, \emptyset\}$, where $\perp$ and $\emptyset$ represent, respectively, a double 
click event, {\it i.e.}, the two detectors click, and a no click event. If the outcome is $\perp$, 
Bob assigns a random bit to the event
\footnote{Note that this random assignment is not mandatory. Indeed, Bob can always choose a particular bit value, say 0, 
  for the double click events.
  This deterministic procedure also preserves the basis-independence detection efficiency condition described in Eq.~(\ref{independentefficiency}).}. 
As a result, Bob obtains $y^{i}\in\{0,1,\emptyset\}$.
The outcomes 0 and 1 will be called a detection event. 
\\\\
    {3. \it Sifting}
    \\
    Bob declares over an authenticated public channel whether or not he obtained a detection event.
    If yes, Alice and Bob annouce their basis choices, and Alice identifies if the event can be assigned to
    the following sets for all $k\in\mathcal{K}$:
    $\mathcal{S}_{Z,Z,k,\U{det}}:=\{i|a^{i}=b^{i}=Z,~k^{i}=k,~y^{i}\neq\emptyset\}$.
    Moreover, if $a^{i}\neq Z$ or $b^{i}\neq Z$, Alice asks Bob to also announce his measurement outcome, and Alice
    identifies if the event can be assigned to
    the following sets for all $c\in\mathcal{C}$, $k\in\mathcal{K}$, $b\in\mathcal{B}$ and $y\in\{0,1\}$:
    $\mathcal{S}_{c,k,\U{det},y,b}:=\{i|c^{i}=c,k^{i}=k,~b^{i}=b,~y^{i}=y\}$. 
Then, Alice checks if the following termination condition is satisfied for a prefixed $N_{\U{det}}$: 
$S_{\U{det}}:=|\mathcal{S}_{\U{det}}|\geq N_{\U{det}}$ for the set $\mathcal{S}_{\U{det}}=\{i|y^{i}\neq\emptyset\}$. 
Once this termination condition is met after sending $N_{\U{sent}}$ pulses, the results associated to the set
$\mathcal{S}_{Z,Z,\U{det}}:=\cup_{k\in\mathcal{K}}\mathcal{S}_{Z,Z,k,\U{det}}$ 
form Alice and Bob's sifted keys $\bm{\kappa}^{\U{sift}}_{A}$ and $\bm{\kappa}^{\U{sift}}_{B}$.
That is, the length of these sifted keys is $|\bm{\kappa}^{\U{sift}}_{A}|=|\bm{\kappa}^{\U{sift}}_{B}|=|\mathcal{S}_{Z,Z,\U{det}}|$.
If the termination condition is not met after sending $N$ pulses, then Alice and Bob abort the protocol. 
\\\\
    {4. \it Parameter estimation}
    \\
    Alice calculates a lower bound for the parameter $S_{Z,Z,n=1,u,\U{det}}:=|\mathcal{S}_{Z,Z,n=1,u,\U{det}}|$,
    where $\mathcal{S}_{Z,Z,n,u,\U{det}}:=\{i|a^{i}=b^{i}=Z,n^{i}=n,t^{i}=u,y^{i}\neq\emptyset\}$
    is a subset of $\mathcal{S}_{Z,Z,\U{det}}$ composed of those elements where Alice emitted an untagged $n$-photon
    state. We call a lower bound on $S_{Z,Z,1,u,\U{det}}$ as $S^{\U{L}}_{Z,Z,1,u,\U{det}}$, which is given by Eq.~(\ref{resultZ1L}).
    Also, she calculates an upper-bound $N^{\U{U}}_{\U{ph},Z,Z,1,u,\U{det}}$ on the number of phase errors
    $N_{\U{ph},Z,Z,1,u,\U{det}}$ for the set $\mathcal{S}_{Z,Z,1,u,\U{det}}$,
    whose quantity is given by Eq.~(\ref{main:Nph}). If the upper bound
    $e^{\U{U}}_{\U{ph}|Z,Z,1,u,\U{det}}:=N^{\U{U}}_{\U{ph},Z,Z,1,u,\U{det}}/S^{\U{L}}_{Z,Z,1,u,\U{det}}$ on the 
    phase error rate
    $e_{\U{ph}|Z,Z,1,u,\U{det}}:=N_{\U{ph},Z,Z,1,u,\U{det}}/S_{Z,Z,1,u,\U{det}}$ satisfies
    $e^{\U{U}}_{\U{ph}|Z,Z,1,u,\U{det}}\geq \overline{e^{\U{U}}_{\U{ph}|Z,Z,1,u,\U{det}}}$, where
    $\overline{e^{\U{U}}_{\U{ph}|Z,Z,1,u,\U{det}}}$ corresponds to the phase error rate associated with a zero secret key rate
    [see Eq.~(\ref{eq:keylength})], Alice and Bob abort the protocol. Otherwise, they proceed to step 5.
\\\\
    {5. \it Bit error correction}
    \\
    Through public discussions, Bob corrects his sifted key $\bm{\kappa}^{\U{sift}}_{B}$ to make it coincide with Alice's key $\bm{\kappa}^{\U{sift}}_{A}$ and obtains
    $\bm{\kappa}^{\U{cor}}_B$ ($|\bm{\kappa}^{\U{cor}}_B|=|\mathcal{S}_{Z,Z,\U{det}}|$).
    \\\\
      {6. \it Privacy amplification}
      \\
      Alice and Bob conduct privacy amplification 
      by shortening $\bm{\kappa}^{\U{sift}}_{A}$ and  $\bm{\kappa}^{\U{cor}}_B$ to obtain the final keys $\bm{\kappa}^{\U{fin}}_A$ and $\bm{\kappa}^{\U{fin}}_B$ of size 
      $|\bm{\kappa}^{\U{fin}}_A|=|\bm{\kappa}^{\U{fin}}_B|=\ell$ with $\ell$ given by Eq.~(\ref{eq:keylength}).

      \section{Secret key generation length}
      \label{Secret key generation length}
      In this section, we present a formula to compute the secret key generation length $\ell$ that guarantees that the protocol
      introduced above is $\epsilon_{\U{sec}}$-secure.
      According to the universal composable security framework~\cite{composable2004,Renner2005},
      we say that a protocol is $\epsilon_{\U{sec}}$-secure if it is both $\epsilon_{\U{c}}$-correct and $\epsilon_{\U{s}}$-secret where
      $\epsilon_{\U{sec}}=\epsilon_{\U{c}}+\epsilon_{\U{s}}$~\cite{Koashi2009}.
      We say that the protocol is $\epsilon_{\U{c}}$-correct if $p(\bm{\kappa}^{\U{fin}}_A\neq \bm{\kappa}^{\U{fin}}_B)\le \epsilon_{\U{c}}$ holds.
      Also, we say that the protocol is $\epsilon_{\U{s}}$-secret if
      \begin{align}
\frac{1}{2}||\hat{\rho}^{\U{fin}}_{AE}-\hat{\rho}^{\U{ideal}}_{AE}||\le \epsilon_{\U{s}}
        \end{align}
      holds in terms of the trace norm, 
      where $\hat{\rho}^{\U{fin}}_{AE}=\sum_{\bm{\kappa}^{\U{fin}}_A}p(\bm{\kappa}^{\U{fin}}_A)\ket{\bm{\kappa}^{\U{fin}}_A}\bra{\bm{\kappa}^{\U{fin}}_A}\otimes\hat{\rho}_E(\bm{\kappa}^{\U{fin}}_A)$ is a classical-quantum state
      between Alice's final key and Eve's system after
      finishing the protocol and $\hat{\rho}^{\U{ideal}}_{AE}$ is an ideal state in which Alice's key is uniformly distributed over
      $2^{|\bm{\kappa}^{\U{fin}}_A|}$ values and decoupled from Eve's system. 
      We suppose that the following two conditions for the random variables $S_{Z,Z,1,u,\U{det}}$ and $N_{\U{ph},Z,Z,1,u,\U{det}}$ are satisfied
      \begin{align}
        &p(S_{Z,Z,1,u,\U{det}}<S^{\U{L}}_{Z,Z,1,u,\U{det}}|S_{\U{det}}=N_{\U{det}})\le \epsilon_{Z},\\
        &p(N_{\U{ph},Z,Z,1,u,\U{det}}>N^{\U{U}}_{\U{ph},Z,Z,1,u,\U{det}}|n_{\U{tag}}\le N_{\U{tag}},S_{\U{det}}=N_{\U{det}})
        \le \epsilon_{\U{PH}}\label{failurePH}
        \end{align}
      regardless of Eve's attack. In this case, for any $\epsilon_{\U{PA}}>0$, by setting~\cite{1367-2630-16-6-063009,PhDKawakami}
      \begin{align}
        \epsilon_{\U{s}}=\sqrt{2}{\sqrt{\epsilon_{\U{PA}}+\epsilon_{\U{PH}}}}+\epsilon_{Z},
        \label{secrecy}
        \end{align}
      it can be shown that the protocol is $\epsilon_{\U{c}}$-correct and $\epsilon_{\U{s}}$-secret if the final key length $\ell$ satisfies
\begin{align}
  \ell\le \ell_{\U{SCIC}}:=S^{\U{L}}_{Z,Z,1,u,\U{det}}\left[1-h\left(\frac{N^{\U{U}}_{\U{ph},Z,Z,1,u,\U{det}}}{S^{\U{L}}_{Z,Z,1,u,\U{det}}}\right)\right]-\log_2\frac{2}{\epsilon_{\U{PA}}}-\lambda_{\U{EC}}(\epsilon_{\U{c}}),
  \label{eq:keylength}
\end{align}
where $h(x)$ is the binary entropy function, and $\lambda_{\U{EC}}(\epsilon_{\U{c}})$ is the cost of error correction to achieve
$\epsilon_{\U{c}}$-correctness. 

\section{results of Parameter estimation}
\label{Parameter estimations}
In this section, we summarise the estimation results of $S^{\U{L}}_{Z,Z,1,u,\U{det}}$ and $N^{\U{U}}_{\U{ph},Z,Z,1,u,\U{det}}$. 
All the detailed derivations of these quantities can be found in Appendices~\ref{sec:appC} and \ref{sec:appD}, respectively. 

First, regarding the estimation of $S^{\U{L}}_{Z,Z,1,u,\U{det}}$, we employ the decoy-state method and we obtain the lower bound on
$S_{Z,Z,1,u,\U{det}}$ as
\begin{align}
  &S^{\U{L}}_{Z,Z,1,u,\U{det}}\notag\\
  &=\frac{\mu^-_{k_1}\sum_{k\in\mathcal{K}}p_k\mu^-_ke^{-\mu^-_k}}{(\mu^+_{k_2}-\mu^-_{k_3})(\mu^-_{k_1}-\mu^+_{k_2}-\mu^-_{k_3})}
\Big\{
  \frac{e^{\mu^-_{k_2}}[S^-_{Z,Z,k_2,u,\U{det}}-g_{\U{MA}}(\epsilon^{Z,k_2,u}_{\U{MA}},p^{B}_Z,N_{\U{det}})]}{p_{k_2}}
-\frac{e^{\mu^+_{k_3}}[S_{Z,Z,k_3,\U{det}}+g_{\U{MA}}(\epsilon^{Z,k_3,u}_{\U{MA}},p^{B}_Z,N_{\U{det}})]}{p_{k_3}}\notag\\
&-\frac{(\mu^+_{k_2})^2-(\mu^-_{k_3})^2}{(\mu^-_{k_1})^2}\Big(\frac{e^{\mu^+_{k_1}}
  [S_{Z,Z,k_1,\U{det}}+g_{\U{MA}}(\epsilon^{Z,k_1,u}_{\U{MA}},p^{B}_Z,N_{\U{det}})]}{p_{k_1}}
                \Big)\Big\}+g_{\U{MA}}(\epsilon^{Z,1,u}_{\U{MA}},p^{B}_Z,N_{\U{det}})
  \label{resultZ1L}
\end{align}
except for error probability $\epsilon_Z:=\sum_{k\in\mathcal{K}}\epsilon^{Z,k,u}_{\U{MA}}+\epsilon^{Z,1,u}_{\U{MA}}+p_{\U{fail}}$ for any
$\epsilon^{Z,k,u}_{\U{MA}}>0$ and $\epsilon^{Z,1,u}_{\U{MA}}>0$.
Here, we define $S^-_{Z,Z,k_2,u,\U{det}}:=S_{Z,Z,k_2,\U{det}}-N_{\U{tag}}$ and the statistical fluctuation term in the Modified Azuma's
inequality (see Appendix~\ref{sec:appAzuma}) is given by 
$g_{\U{MA}}(\epsilon,q,n)=\frac{\sqrt{\ln\epsilon(\ln\epsilon-18nq)}-\ln\epsilon}{3}$. 

Second, in the estimation of the number of phase errors $N_{\U{ph},Z,Z,1,u,\U{det}}$ 
for the untagged single-photon emission events in $\bm{\kappa}^{\U{sift}}_{A}$, we follow
the arguments in the loss-tolerant protocol in~\cite{PhysRevA.93.042325,AkihiroNJP2015,PhysRevA.90.052314}.
In the main text, for simplicity of its expression, we only describe $N_{\U{ph},Z,Z,1,u,\U{det}}$ with the following restricted phase intervals:
\begin{align}
R^{0_Z}_{\U{\U{ph}}}=[-\theta,\theta],~~
R^{1_Z}_{\U{\U{ph}}}=[\pi-\theta,\pi+\theta],~~
R^{0_X}_{\U{\U{ph}}}=\left[\frac{\pi}{2}-\theta,\frac{\pi}{2}+\theta\right]~~~\left(\U{with}~0\le\theta<\frac{\pi}{6}\right).
\label{simpleRph}
\end{align}
Note that the expression of $N_{\U{ph},Z,Z,1,u,\U{det}}$ with the general phase intervals in Eq.~(\ref{Rph}) 
  should be refereed to Appendix~\ref{apEgeneralRph}. 
  Under the assumption of Eq.~(\ref{simpleRph}), $N^{\U{U}}_{\U{ph},Z,Z,1,u,\U{det}}$ can be written as a linear combination of 
the parameters $S'_{c,1,u,\U{det},y,X}$, which are bounds on the cardinality of the sets 
$\mathcal{S}_{c,1,u,\U{det},y,X}=\{i|c^{i}=c,n^{i}=1,t^i=u,~b^{i}=X,~y^{i}=y)\}$, as
\begin{align}
  N^{\U{U}}_{\U{ph},Z,Z,1,u,\U{det}}=\frac{p^A_Zp^B_Z(1+\sin\theta)}{2}\sum^1_{y=0}\sum_{c\in\mathcal{C}}\Gamma_{y,c}^{\U{U}}
  \frac{S'_{c,1,u,\U{det},y\oplus1,X}+\U{sgn}(\Gamma_{y,c}^{\U{U}})g_{\U{A}}(N_{\U{det}},\epsilon^{c,1,u,y,X}_{\U{A}})}
       {p(c)p^B_X}+g_{\U{A}}(N_{\U{det}},\epsilon^{\U{ph},Z,1,u}_{\U{A}}).
    \label{main:Nph}
\end{align}
Here, we define the statistical fluctuation term of the Azuma's inequality~\cite{Azuma1967} as
$g_{\U{A}}(x,y):=\sqrt{2x\ln1/y}$ and the functions $\{\Gamma_{y, c}^{\U{U}}\}_{y,c}$~\cite{PhysRevA.93.042325} as
$\Gamma_{0,0_Z}^{\U{U}}=\frac{\sin{\theta}}{\sin{\theta}+\cos{\frac{3}{2}\theta}}$ $(0\le\Gamma_{0,0_Z}^{\U{U}}<\sqrt{2}-1)$,
$\Gamma_{0,1_Z}^{\U{U}}=\Gamma_{0,0_Z}^{\U{U}}$, 
$\Gamma_{0,0_X}^{\U{U}}=\frac{1-\sin{\theta}}{\cos{2\theta}-\sin{\theta}}$ $(1\le\Gamma_{0,0_X}^{\U{U}}<\infty)$, 
$\Gamma_{1,0_Z}^{\U{U}}=\frac{\cos{\theta}}{\cos{\theta}-\sin{\frac{3}{2}\theta}}$ $(1\le\Gamma_{1,0_Z}^{\U{U}}<3+\sqrt{6})$,
$\Gamma_{1,1_Z}^{\U{U}}=\Gamma_{1,0_Z}^{\U{U}}$ and
$\Gamma_{1,0_X}^{\U{U}}=-\frac{1-\sin{\theta}}{1+\sin{\theta}}$ $(-1\le\Gamma_{1,0_X}^{\U{U}}<-1/3)$. 
Regarding $S'_{c,1,u,\U{det},y,X}$, we take the following upper or lower bounds on $S_{c,1,u,\U{det},y,X}:=|\mathcal{S}_{c,1,u,\U{det},y,X}|$, 
depending on the sign of $\Gamma^{\U{U}}_{y,c}$, such that $N_{\U{ph},Z,Z,1,u,\U{det}}$ takes its upper bound:
\begin{align}
S'_{c,1,u,\U{det},y,X}=
  \begin{cases}
  S^{\U{U}}_{c,1,u,\U{det},y,X}~~~\U{if}~~\Gamma^{\U{U}}_{y,c}>0,\\
  S^{\U{L}}_{c,1,u,\U{det},y,X}~~~\U{if}~~\Gamma^{\U{U}}_{y,c}\le0,
\end{cases}
\end{align}
with
\begin{align}
S^{\U{U}}_{c,1,u,\U{det},y,X}:&=
  \frac{\frac{[S_{c,k_2,\U{det},y,X}+g_{\U{MA}}(\epsilon^{c,k_2,u,y,X}_{\U{MA}},p^{B}_X,N_{\U{det}})]e^{-\mu^-_{k_3}}}{p_{k_2}}-
    \frac{[S^-_{c,k_3,u,\U{det},y,X}-g_{\U{MA}}(\epsilon^{c,k_3,u,y,X}_{\U{MA}},p^{B}_X,N_{\U{det}})]e^{-\mu^+_{k_2}}}{p_{k_3}}}
       {e^{-\mu^-_{k_2}-\mu^-_{k_3}}\mu^-_{k_2}-e^{-\mu^+_{k_2}-\mu^+_{k_3}}\mu^+_{k_3}}\notag\\
       &\times \sum_{k\in\mathcal{K}}p_k\mu^+_ke^{-\mu^+_k}+g_{\U{MA}}(\epsilon^{c,1,u,y,X}_{\U{MA}},p^{B}_X,N_{\U{det}})
       \label{main:singlegain1}
\end{align}
and
\begin{align}
S^{\U{L}}_{c,1,u,\U{det},y,X}:&=
\frac{\mu^-_{k_1}\sum_{k\in\mathcal{K}}p_k\mu^-_ke^{-\mu^-_k}}{(\mu^+_{k_2}-\mu^-_{k_3})(\mu^-_{k_1}-\mu^+_{k_2}-\mu^-_{k_3})}
\Big\{
  \frac{e^{\mu^-_{k_2}}[S^-_{c,k_2,u,\U{det},y,X}-g_{\U{MA}}(\epsilon^{c,k_2,u,y,X}_{\U{MA}},p^{B}_X,N_{\U{det}})]}{p_{k_2}}\notag\\
 & -\frac{e^{\mu^+_{k_3}}[S_{c,k_3,\U{det},y,X}+g_{\U{MA}}(\epsilon^{c,k_3,u,y,X}_{\U{MA}},p^{B}_X,N_{\U{det}})]}{p_{k_3}}\notag\\
&-\frac{(\mu^+_{k_2})^2-(\mu^-_{k_3})^2}{(\mu^-_{k_1})^2}
  \Big(\frac{e^{\mu^+_{k_1}}[S_{c,k_1,\U{det},y,X}+g_{\U{MA}}(\epsilon^{c,k_1,u,y,X}_{\U{MA}},p^{B}_X,N_{\U{det}})]}{p_{k_1}}
  \Big)\Big\}+g_{\U{MA}}(\epsilon^{c,1,u,y,X}_{\U{MA}},p^{B}_X,N_{\U{det}}),
  \label{main:singlegain2}
\end{align}
where $S^-_{c,k,u,\U{det},y,X}:=S_{c,k,\U{det},y,X}-N_{\U{tag}}$. 
Finally, we calculate the failure probability $\epsilon_{\U{PH}}$ associated to the estimation of $N^{\U{U}}_{\U{ph},Z,Z,1,u,\U{det}}$ 
in Eq.~(\ref{failurePH}), as 
$\epsilon_{\U{PH}}=\epsilon^1_{\U{PH}}+\epsilon^{2}_{\U{PH}}$, and we define 
$\epsilon^1_{\U{PH}}:=\sum^1_{y=0}\sum_{c\in\mathcal{C}}\epsilon^{c,1,u,y,X}_{\U{A}}+\epsilon^{\U{ph},Z,1,u}_{\U{A}}$ 
for any $\epsilon^{c,1,u,y,X}_{\U{A}}>0$ and $\epsilon^{\U{ph},Z,1,u}_{\U{A}}>0$. 
$\epsilon^{2}_{\U{PH}}$, on the other hand, is composed of the failure probabilities associated to the estimation of $\{S_{c,1,u,\U{det},y,X}\}_{y=0,1,c\in\mathcal{C}}$, and
$\epsilon^{2}_{\U{PH}}$ has the form $\epsilon^{2}_{\U{PH}}=\sum_{y=0,1}\sum_{c\in\mathcal{C}}\epsilon^{c,1,u,y,X}$
with
$\epsilon^{c,1,u,y,X}=\sum_{k=k_2,k_3}\epsilon^{c,k,u,y,X}_{\U{MA}}+\epsilon^{c,1,u,y,X}_{\U{MA}}$ or
$\epsilon^{c,1,u,y,X}=\sum_{k\in\mathcal{K}}\epsilon^{c,k,u,y,X}_{\U{MA}}+\epsilon^{c,1,u,y,X}_{\U{MA}}$ 
depending on whether we use the upper bound given by Eq.~(\ref{main:singlegain1}) or the lower bound given by Eq.~(\ref{main:singlegain2}).

\section{Simulation of the key rate}
\label{keyrate}
\begin{figure}[t]
\includegraphics[width=8.5cm]{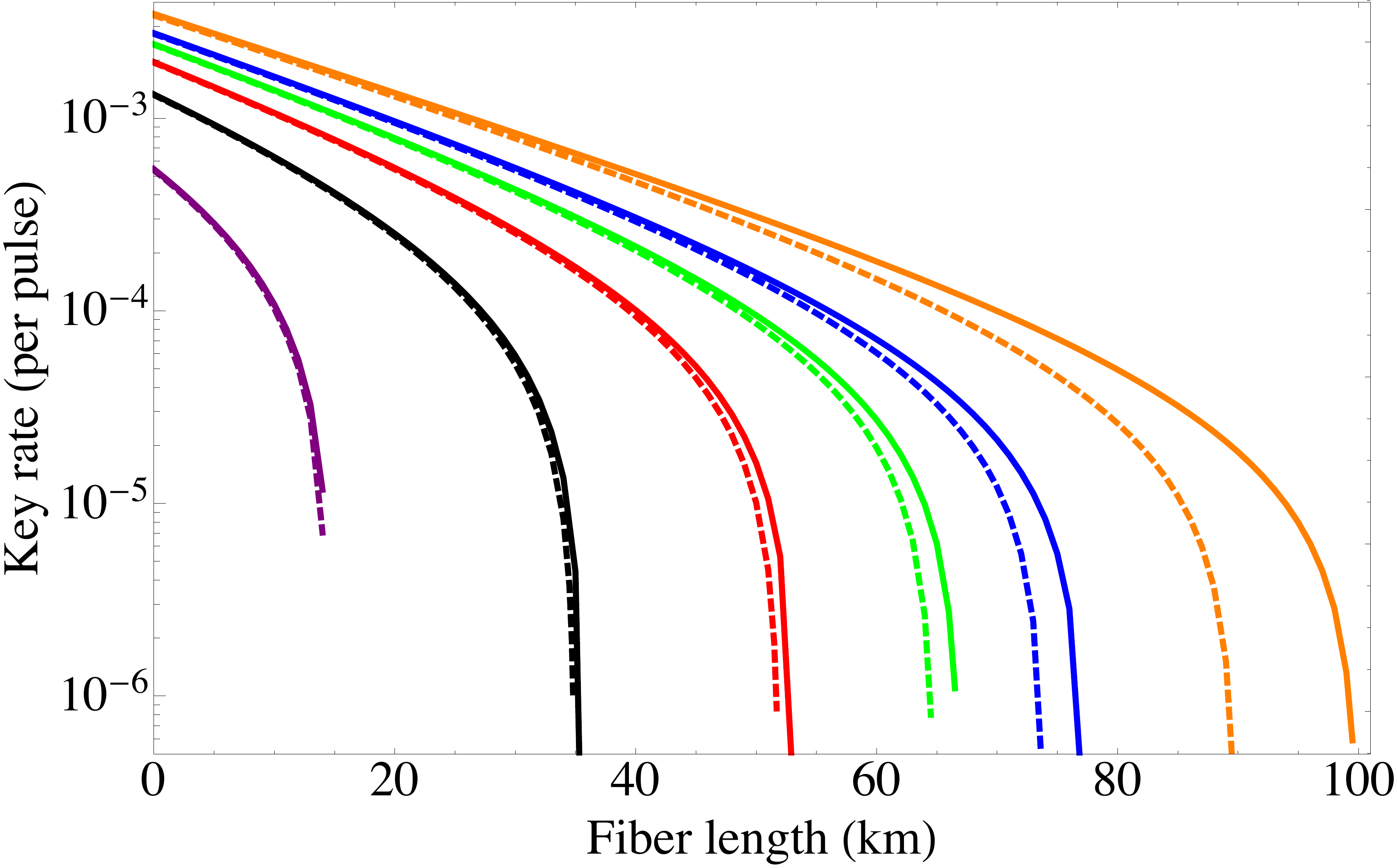}
\caption{
  The key rate (per pulse) in logarithmic scale versus fibre length for the case with the phase fluctuation of $\pm 0.03$~rad
  [namely, $\theta=0.03$ in Eq.~(\ref{simpleRph})]
  for any choice of $c^i\in\mathcal{C}$ and the intensity fluctuation of
  $\pm 3\%$ for the choice of $k^i\in\{k_1,k_2\}$ and $R^{k_3}_{\U{int}}=[0,10^{-3}]$ for the weakest decoy setting $k^i=k_3$.
  In solid lines, we assume (I)~$N_{\U{tag}}$=0 and $p_{\U{fail}}=0$, and in the dashed lines, we assume
    (II)~$N_{\U{tag}}=N_{\U{sent}}\times 10^{-7}$ and $p_{\U{fail}}=0$. 
  The secrecy and correctness parameters are $\epsilon_{\U{s}}=\epsilon_{\U{c}}=10^{-10}$ and for each set of solid and dashed lines, 
  the total number of signals sent by Alice is $N_{\U{sent}}\in\{10^{10},10^{10.5},10^{11},10^{11.5},10^{12}\}$
  from left to right. The rightmost solid and dashed lines respectively correspond to the asymptotic key rate of the cases (I) and
    (II) where no statistical
  flucuation terms in $N^{\U{U}}_{\U{ph},Z,Z,1,u,\U{det}}$ and $S^{\U{L}}_{Z,Z,1,u,\U{det}}$ are taken into account.
  The experimental parameters are described in the main text. 
}
  \label{fig:keyrate}
\end{figure}
In this section, we show the numerical simulation results of the key rate for a fibre-based QKD system.
In the simulation,
we assume that Bob uses a measurement setup with two single-photon detectors with detection efficiency $\eta_{\U{det}}=10\%$
and dark count probability per pulse $p_{\U{dark}}=10^{-5}$. These parameters are set to be the same as those
in~\cite{PhysRevA.96.012305}.
The attenuation coefficient of the optical fibre is $0.2$ dB/km and its transmittance is $\eta_{\U{ch}}=10^{-0.2l/10}$
with $l$ denoting the fibre length.
We denote the channel transmission rate including detection efficiency by $\eta:=\eta_{\U{ch}}\eta_{\U{det}}$. 
The overall misalignment error of the measurement system is fixed to be $e_{\U{mis}}=1\%$.
In addition, we assume an  error correction cost equals to 
$\lambda_{\U{EC}}(\epsilon_{\U{c}})=1.05\times|\bm{\kappa}^{\U{sift}}_{A}|h(e_{\U{bit}})+\log_{2}(1/\epsilon_{\U{c}})$,
where $e_{\U{bit}}$ is the bit error rate of the sifted key $(\bm{\kappa}^{\U{sift}}_{A}, \bm{\kappa}^{\U{sift}}_{B})$.
Moreover, we suppose that the intervals of the intensity fluctuation in Eq.~(\ref{Rint}) are given by 
$R^k_{\U{int}}=[\mu_k(1-r_k),\mu_k(1+r_k)]$ for $k\in\{k_1,k_2\}$ and $R^{k_3}_{\U{int}}=[0,10^{-3}]$ with $\mu_k$ denoting the expected 
intensity (we suppose that $\mu_{k_3}=0$)
and where $r_k$ represents the deviation of the actual intensity from the expected value.
For the intervals of the phase fluctuation $R^{c^i}_{\U{ph}}$ in Eq.~(\ref{simpleRph}), we take the experimental value of phase modulation error from Ref.~\cite{PhysRevA.92.032305}
and we set $\theta=0.03$~rad
\footnote{
  Note that the phase fluctuation $\pm$0.03 rad is the largest
  experimental value observed in~\cite{PhysRevA.92.032305} in a protocol that uses the three states $\{0,\pi/2,3\pi/2\}$
  (see table~III in~\cite{PhysRevA.92.032305}). 
  Note also that the authors of~\cite{PhysRevA.92.032305} measured a different quantity than the one we need here. 
  More precisely, Ref.~\cite{PhysRevA.92.032305} assumes that each sending pulse is independent and identically distributed~(IID) and
  measured the fixed deviation due to imperfect encoding. 
In contrast, we allow each pulse to be different (non-IID case) 
and we are interested in the deviation of each pulse from the mean. In our simulation,
we simply assume that the result in Ref.~\cite{PhysRevA.92.032305} gives us some reasonable estimation on the deviation $\theta$. 
  }. 
In Fig.~\ref{fig:keyrate}, we consider the two cases: (I)~$N_{\U{tag}}=0$ and $p_{\U{fail}}=0$ and 
(II)~$N_{\U{tag}}=N_{\U{sent}}\times 10^{-7}$ and $p_{\U{fail}}=0$. 
The case (I) means that all the phases and the intensities lie in their intervals, and 
the case (II) means that the number of tagged events is upper bounded by $N_{\U{sent}}\times 10^{-7}$. 

Regarding the numbers of detection events $S_{Z,Z,k,\U{det}}$ and $S_{c,k,\U{det},y,b}$, we generate these quantities by assuming the following specific setup. 
In particular, for the $i^{\U{th}}$ trial, we consider that Alice sends Bob pairs of coherent states through the fibre of the form
$\ket{e^{\U{i}(\delta+\theta_c)}\sqrt{\mu_k/2}}_{S^i}\ket{e^{\U{i}\delta}\sqrt{\mu_k/2}}_{R^i}$ with $\theta_{0_Z}=0$, $\theta_{1_Z}=\pi$ and $\theta_{0_X}=\pi/2$
according to the choices of $k^i=k\in\mathcal{K}$ and $c^i=c\in\mathcal{C}$
\footnote{
  Note that this assumption is used just to simulate the experimentally observed numbers
  (namely, $S_{Z,Z,k,\U{det}}$ and $S_{c,k,\U{det},y,b}$), and we do not require this assumption in the actual experiments. 
}.
Bob measures the incoming signals using a Mach-Zehnder interferometer with two 50:50 BSs and a phase modulator
as shown in Fig.~\ref{fig:actual}. 
More precisely, he uses a first 50:50 BS to split the received pulses into two beams, and after that
he applies a phase shift 0 or $\pi/2$ to one of them according to his basis choice of $b^{i}=Z$ or $b^{i}=X$, respectively, and finally
he lets interfere the resulting pulses using a second 50:50 BS. 
In this setup, we obtain the following probabilities: 
$p(y^i=y|c^i=y_Z,k^i=k,b^i=Z)=p(y^i=0|c^i=0_X,k^i=k,b^i=X)=[1-e^{-\frac{\eta\mu_k}{2}}(1-p_{\U{d}})](1-\frac{p_{\U{d}}}{2})$, 
$p(y^i=y\oplus1|c^i=y_Z,k^i=k,b^i=Z)=p(y^i=1|c^i=0_X,k^i=k,b^i=X)=\frac{p_{\U{d}}[1+e^{-\frac{\eta\mu_k}{2}}(1-p_{\U{d}})]}{2}$ 
for $y\in\{0,1\}$,
and $p(y^i=y|c^i=x_Z,k^i=k,b^i=X)=p(y^i=y|c^i=0_X,k^i=k,b^i=Z)=\frac{1-(1-p_{\U{d}})^2e^{-\frac{\eta\mu_k}{2}}}{2}$
for $y,x\in\{0,1\}$.
Moreover, we assume that the bit error rate $e_{\U{bit}}$ is given by 
$e_{\U{bit}}=\sum_{y=0,1}p(y^i=y\oplus1|c^i=y_Z,k^i=k_1,b^i=Z)/\sum_{x,y=0,1}p(y^i=y|c^i=x_Z,k^i=k_1,b^i=Z)+e_{\U{mis}}$. 
With these probabilities, we suppose that the experimentally observed numbers satisfy 
$S_{Z,Z,k,\U{det}}={N_{\U{sent}}}\sum_{x,y=0,1}\frac{p^A_Z}{2}p^B_Zp_k\times p(y^i=y|c^i=x_Z,k^i=k,b^i=Z)$ and 
$S_{c,k,\U{det},y,b}={N_{\U{sent}}}p(c)p^B_bp_k\times p(y^i=y|c^i=c,k^i=k,b^i=b)$. 

With the above parameters, we simulate the key rate $\ell_{\U{SCIC}}/{N_{\U{sent}}}$ for a fixed value of the correctness and
secrecy parameters $\epsilon_{\U{c}}=\epsilon_{\U{s}}=10^{-10}$ and we set $\epsilon_Z=1/2\times 10^{-10}$,
$\epsilon_{\U{PA}}=\epsilon_{\U{PH}}=1/16\times 10^{-20}$, and 
$\epsilon^{Z,k,u}_{\U{MA}}=\epsilon^{Z,1,u}_{\U{MA}}=\epsilon_Z/4=1/8\times 10^{-10}$.
We also assume that each failure probability which is contained in the expression of $\epsilon^1_{\U{PH}}$ and $\epsilon^2_{\U{PH}}$
takes the value 
$\epsilon^{c,1,u,y,X}_{\U{A}}=\epsilon^{\U{ph},Z,1,u}_{\U{A}}=1/26\times\epsilon_{\U{PH}}$ and $\epsilon^{c,k,u,y,X}_{\U{MA}}=
\epsilon^{c,1,u,y,X}_{\U{MA}}=1/26\times\epsilon_{\U{PH}}$, respectively, and we set 
$p^A_Z=p^B_Z=p_{k_1}=0.8$ and $p_{k_2}=0.1$. 
In the simulation, we perform a numerical optimization of the key rate $\ell_{\U{SCIC}}/{N_{\U{sent}}}$ over the two free parameters $\mu_{k_1}$ and $\mu_{k_2}$.
  In the solid and dashed lines in Fig.~\ref{fig:keyrate}, we respectively plot the key rate of the cases (I) and (II) for the finite-case when $N_{\U{sent}}\in\{10^{10},10^{10.5},10^{11},10^{11.5},10^{12}$\} (from left to right). 
  For comparison, the rightmost solid and dashed lines respectively correspond to the asymptotic key rate of the cases (I) and (II), 
  where no statistical fluctuation terms in Eqs.~(\ref{resultZ1L}) and (\ref{main:Nph}) are taken into account.
  Our simulation results show the feasibility of secure key distribution within a reasonable time by employing practical devices
  that satisfy our device assumptions. For instance, if Alice uses a laser diode operating at 1~GHz repetition rate and she sends
  $N_{\U{sent}}=10^{12}$ signals, then we find that it is possible to distribute a 1-Mb secret key over a 75-km fibre link in $<0.3$ hours.
  This scenario corresponds to the solid blue line (the fifth solid line from the left) shown in Fig.~\ref{fig:keyrate}.

\section{Conclusion}
\label{Conclusion}
In summary, we have provided an information theoretic
security proof for the loss-tolerant QKD protocol which accommodates the setting-choice-independent
correlation (SCIC) in the finite key regime. 
Within the framework of SCIC, the relative phases and intensities of the sending coherent states fluctuate over time. 
Once realistic intervals for these fluctuations 
(such as for instance $\pm 0.03$~rad and $\pm 3\%$, respectively) are guaranteed, our numerical simulations have shown that secure quantum communication is feasible with a reasonable number of signal transmissions such as for example $N_{\U{sent}}=10^{12}$. 
Therefore, our results constitute a significant step toward realising secure quantum communication with practical source devices. 
On a more general outlook, it would be of great practical interest to improve the convergence speed of the key rate in the 
finite-key regime and it would be also important to weaken the requirements for Alice's source device which could
lead to a simpler characterisation of the source devices.

\section*{Acknowledgements}
\label{Acknowledgments}
We thank Masato Koashi, Toshihiko Sasaki and Tatsuya Sumiya for crucial comments on the correlation model of the source devices, and 
Yuki Takeuchi for valuable discussions on the security analysis. 
AM acknowledges support from Grant-in-Aid for JSPS Fellows (KAKENHI Grant No. JP17J04177).
GK, KA, and KT acknowledge support from ImPACT Program of Council for Science, Technology and Innovation (Cabinet Office, Government of Japan).
MC acknowledges support from the Spanish Ministry of Economy and Competitiveness (MINECO),
the Fondo Europeo de Desarrollo Regional (FEDER) through grant TEC2014-54898-R, and the European Commission (project “QCALL”).
NI acknowledges support from JST-CREST JPMJCR 1671. HKL acknowledges financial support from the Natural Sciences and Engineering
Research Council of Canada (NSERC), the US Office of Naval Research (ONR), 
Canadian Foundation for Innovation (CFI), Ontario Research Fund (ORF) and Post-secondary Strategic Infrastructure Fund (SIF).
\appendix

\section{Summary of notations}
A list of the parameters used throughout this paper together with their notation is provided in Tables~\ref{table:1} and \ref{table:2}. 

\begin{table}[h]
\begin{center}
\begin{tabular}
{|c|c|}\hline 
Random variables and sets &Meaning\\ \hline\hline
$\bm{X}^{i}$ &Abbreviation of $X^1,X^{2},...,X^i$ \\
&($X^i$ can be $c^i,k^i,\theta^i,\mu^i,P^i,g^i,n^i,y^i,b^i$)\\\hline
$\mathcal{C}$ & Set of bit and basis information: $\{0_Z,1_Z,0_X\}$\\ \hline
$c^i\in\mathcal{C}$ & Alice's $i^{\U{th}}$ choice of bit and basis information\\ \hline
$\mathcal{K}$ & Set of intensity setting: $\{k_1,k_2,k_3\}$\\ \hline
$k^i\in\mathcal{K}$ & Alice's $i^{\U{th}}$ choice of intensity setting\\ \hline
$\theta^i\in[0,2\pi)$ &$i^{\U{th}}$ relative phase between the signal and reference pulses \\ \hline
$\mu^i\in[0,\infty)$ &$i^{\U{th}}$ total intensity in both the signal and reference pulses \\ \hline
    $\mathcal{G}^i_{\U{unt}}$ &Untagged set of the $i^{\U{th}}$ internal state of the source device\\ \hline
    $g^i\in\mathcal{G}^i_{\U{unt}}\cup\overline{\mathcal{G}^i_{\U{unt}}}$ &$i^{\U{th}}$ internal state of Alice's source device \\ \hline
    $\bm{g}^{\ge i}$ &Abbreviation of $g^i,g^{i+1},...,g^{N_{\U{sent}}}$ \\\hline
    $t^i\in\{u,t\}$ &Indicator of $g^i\in\mathcal{G}^i_{\U{unt}}$ or $g^i\notin\mathcal{G}^i_{\U{unt}}$: 
      $t^i=u$ if $g^i\in\mathcal{G}^i_{\U{unt}}$ and $t^i=t$ if $g^i\notin\mathcal{G}^i_{\U{unt}}$
    \\\hline
    $n_{\U{tag}}$ &Number of tagged signals: $|\{i|t^i=t\}|$\\ \hline
    $\bm{\mathcal{G}}^{N_{\U{sent}}}_{\U{good}}$ &Set of $\bm{g}^{N_{\U{sent}}}$ where $n_{\U{tag}}$ is upper bounded by a fixed number\\ \hline
    $n^i\in[0,\infty)$ &Total number of photons in the $i^{\U{th}}$ signal and reference pulses \\ \hline
$\mathcal{B}$ &Set of bases: $\{Z,X\}$\\ \hline
      $a^i (b^i)\in\mathcal{B}$ &Alice's (Bob's) $i^{\U{th}}$ basis choice\\ \hline
      $y^i\in\{0,1,\emptyset\}$ &Bob's $i^{\U{th}}$ measurement outcome\\ \hline
      $\mathcal{S}_{Z,Z,k,\U{det}}$ &Set $\{i|a^{i}=b^{i}=Z,~k^{i}=k,~y^{i}\neq\emptyset\}$\\ \hline
$\mathcal{S}_{Z,Z,\U{det}}$ &Set $\{i|a^{i}=b^{i}=Z,~y^{i}\neq\emptyset\}$\\ \hline
$\mathcal{S}_{Z,Z,n,u,\U{det}}$ &Set $\{i|a^{i}=b^{i}=Z,~n^{i}=n,~t^{i}=u,y^i\neq\emptyset\}$\\ \hline
$\mathcal{S}_{c,k,\U{det},y,b}$ &Set $\{i|c^{i}=c,k^i=k,b^{i}=b,~y^i=y\}$\\ \hline
$\mathcal{S}_{c,n,u,\U{det},y,b}$ &Set $\{i|c^{i}=c,n^i=n,t^{i}=u,b^{i}=b,~y^i=y\}$\\ \hline
      $\mathcal{S}_{\U{det}}$ &Set $\{i|y^{i}\neq\emptyset\}$\\ \hline
      $\mathcal{S}^{\U{L}}_{Z,Z,1,u,\U{det}}$ & Lower bound on $|\mathcal{S}_{Z,Z,1,u,\U{det}}|$\\ \hline
      $N_{\U{ph},Z,Z,1,u,\U{det}}$   & Number of phase errors in the set $\mathcal{S}_{Z,Z,1,u,\U{det}}$\\ \hline
      $N^{\U{U}}_{\U{ph},Z,Z,1,u,\U{det}}$   & Upper bound on $N_{\U{ph},Z,Z,1,u,\U{det}}$ \\ \hline
      \end{tabular}
\end{center}
\caption{
Random variables and sets used throughout the paper
}
 \label{table:1}
\end{table}

\begin{table}[h]
\begin{center}
\begin{tabular}
{|c|c|}\hline 
      Quantum systems&Meaning\\ \hline\hline
$C^i$& Quantum system storing $c^i$ possessed by Alice\\ \hline
$K^i$& Quantum system storing $k^i$ possessed by Alice\\ \hline
$\Theta^i$& Quantum system storing $\theta^i$ possessed by Alice\\ \hline
$M^i$& Quantum system storing $\mu^i$ possessed by Alice\\ \hline
$G^i$& Quantum system storing $g^i$ possessed by Alice\\ \hline
$N^i$& Quantum system storing $n^i$ possessed by Alice\\ \hline
$B^i$& $i^{\U{th}}$ quantum system sent to Bob\\ \hline
$A^i$& Alice's $i^{\U{th}}$ quantum system defined iff $t^i=u$ and $n^i=1$\\ \hline
      $B'^i$& Bob's $i^{\U{th}}$ quantum system after Eve's intervention\\ \hline\hline
      Symbols &Meaning\\ \hline\hline
$N_{\U{sent}}$ & Number of pulse pairs (signal and reference pulses) sent by Alice\\ \hline
  $R^{c^i}_{\U{ph}}$ &$i^{\U{th}}$ interval of the relative phase $\theta^i$ for $c^i$\\ \hline
  $R^{k^i}_{\U{int}}$ &$i^{\U{th}}$ interval of the intensity $\mu^i$ for $k^i$\\ \hline
$N_{\U{tag}}$ &Predetermined upper bound on $n_{\U{tag}}$\\ \hline
$p_{\U{fail}}$ &Upper bound on the probability that $\bm{g}^{N_{\U{sent}}}\notin\bm{\mathcal{G}}^{N_{\U{sent}}}_{\U{good}}$ occurs\\ \hline
$N_{\U{det}}$ &Predetermined number of detection events $|\mathcal{S}_{\U{det}}|$\\ \hline
$\bm{\kappa}^{\U{sift}}_{A(B)}$ &Alice's (Bob's) sifted key\\ \hline
$\bm{\kappa}^{\U{cor}}_{B}$ &Bob's reconciled key\\ \hline
$\bm{\kappa}^{\U{fin}}_{A(B)}$ &Alice's (Bob's) final key\\ \hline
$\epsilon_{\U{sec}}$ &Security parameter of the protocol\\ \hline
$\epsilon_{\U{s}}$ &Secrecy parameter of the protocol\\ \hline
$\epsilon_{\U{c}}$ &Correctness parameter of the protocol\\ \hline
      $\epsilon_{\U{PA}}$ &Error probability of privacy amplification\\ \hline
      $\epsilon_Z$ &Failure probability of the estimation of $\mathcal{S}^{\U{L}}_{Z,Z,1,u,\U{det}}$ given that $S_{\U{det}}=N_{\U{det}}$
      \\ \hline
      $\epsilon_{\U{PH}}$ &Failure probability of the estimation of $N^{\U{U}}_{\U{ph},Z,Z,1,u,\U{det}}$ given that $n_{\U{tag}}\le N_{\U{tag}}$ and
      $S_{\U{det}}=N_{\U{det}}$\\ \hline
      $\ell$ &Length (in bits) of the final key\\ \hline

\end{tabular}
\end{center}
\caption{
Quantum systems and symbols used throughout the paper
}
 \label{table:2}
\end{table}

  \section{Description of Alice's sending state}
  \label{Proofofsendingstate}
  For later discussions, we introduce the following theorem about the description of Alice's sending states,
  which is a direct consequence of the assumptions (A-1)-(A-5) in Sec.~\ref{Assumptions on the devices}. 
     \begin{theorem}
       From the assumptions (A-1)-(A-5), we have that all the $N_{\U{sent}}$ sending states can be written as
       \begin{align}
  \sum_{\bm{g}^{N_{\U{sent}}}}\sqrt{p(\bm{g}^{N_{\U{sent}}})}\ket{\bm{g}^{N_{\U{sent}}}}_{\bm{G}^{N_{\U{sent}}}}
  \bigotimes^{N_{\U{sent}}}_{i=1}\sum_{c^{i},k^{i}}\sqrt{p(c^{i})p(k^{i})}\ket{\hat{\rho}^{i}(\theta^i_{c^i,g^{i}},\mu^i_{k^i,g^{i}})}_{N^iB^i}
  \ket{c^{i},k^{i}}_{C^iK^i}\ket{\theta^i_{c^i,g^{i}},\mu^i_{k^i,g^{i}}}_{\Theta^i,M^i}
                   \label{sendingrho}
          \end{align}
       with
       $\bm{G}^{{N_{\U{sent}}}}:=G^1,...,G^{N_{\U{sent}}}$. Here, $G^i$, $C^i$, $K^i$, $\Theta^i$ and $M^i$ denote the systems storing the
       information of $g^i$, $c^i$, $k^i$, $\theta^i$ and $\mu^i$, respectively. According to the assumption (A-1), note that these
       five systems are possessed by Alice.
                      This is to ensure that there is no side channel leaking information to Eve. 
                      Also, $\ket{\hat{\rho}^{i}(\theta^i_{c^i,g^{i}},\mu^i_{k^i,g^{i}})}_{N^iB^i}$ is a purified 
                      state of Eq.~(\ref{phaserandom2}) with $N^i$ being a system storing the number of photons contained in the
                      $i^{\U{th}}$ sending state, which has the form
\begin{align}
\ket{\hat{\rho}^{i}(\theta^i_{c^i,g^{i}},\mu^i_{k^i,g^{i}})}_{N^iB^i}=
   \sum_{n^{i}}\sqrt{p(n^{i}|\mu^{i}_{k^{i},g^i})}\ket{n^{i}}_{N^i}\ket{\hat{\Upsilon}^{i}(\theta^{i}_{c^{i},g^i},n^{i})}_{B^i}. 
  \end{align}
                      \label{th1}
     \end{theorem}
         {\bf Proof of Theorem~\ref{th1}}~~For the most general case, the total $N_{\U{sent}}$ sending states can be coherently written as
        \begin{align}
          \sum_{\bm{g}^{{N_{\U{sent}}}},\bm{\theta}^{{N_{\U{sent}}}},\bm{\mu}^{{N_{\U{sent}}}},\bm{c}^{{N_{\U{sent}}}},\bm{k}^{{N_{\U{sent}}}}}\sqrt{p(\bm{g}^{{N_{\U{sent}}}},\bm{P}^{{N_{\U{sent}}}})}
          \bigotimes^{N_{\U{sent}}}_{i=1}  \ket{\hat{\rho}^{i}(\theta^{i},\mu^{i})}_{N^i,B^i}
          \ket{\theta^{i}}_{\Theta^i}
          \ket{\mu^{i}}_{M^i}\ket{c^{i}}_{C^i}\ket{k^{i}}_{K^i}\ket{g^{i}}_{G^i}.
          \label{generalstate}
        \end{align}
        Next, we calculate $p(\bm{g}^{{N_{\U{sent}}}},\bm{P}^{{N_{\U{sent}}}})$ using the assumptions (A-2)-(A-5).
         \begin{align}
           p(\bm{g}^{{N_{\U{sent}}}},\bm{P}^{{N_{\U{sent}}}})&=\Pi^{N_{\U{sent}}}_{i=1}p(g^i,P^i|\bm{g}^{i-1},\bm{P}^{i-1})\\
           &=\Pi^{N_{\U{sent}}}_{i=1}p(P^i|\bm{g}^{i},\bm{P}^{i-1})p(g^i|\bm{g}^{i-1},\bm{P}^{i-1})\\
           &=\Pi^{N_{\U{sent}}}_{i=1}p(P^i|\bm{g}^{i},\bm{P}^{i-1})p(g^i|\bm{g}^{i-1})\\
           &=\Pi^{N_{\U{sent}}}_{i=1}p(\theta^i,\mu^i|c^i,k^i,\bm{g}^{i},\bm{P}^{i-1})p(c^i,k^i|\bm{g}^{i},\bm{P}^{i-1})p(g^i|\bm{g}^{i-1})\\
           &=\Pi^{N_{\U{sent}}}_{i=1}\delta(\theta^i,\theta^i_{c^i,g^{i}})\delta(\mu^i,\mu^i_{k^i,g^{i}})p(c^i)p(k^i)p(g^i|\bm{g}^{i-1}),
         \end{align}
where in the 3$^{\U{rd}}$ equation we use the assumption (A-2), and in the 5$^{\U{th}}$ equation we use the assumptions (A-3)-(A-5). 
With this expression, Eq.~(\ref{generalstate}) results in Eq.~(\ref{sendingrho}), which ends the proof.\sq

\section{Modified Azuma's inequality}
\label{sec:appAzuma}
Here, we introduce the Modified Azuma's inequality~\cite{Azuma2002} that we use in the decoy-state method (see Appendix~\ref{sec:appC}). 
\begin{theorem}
  \label{LemmaAzumaGeneral}(Modified Azuma's inequality)~\cite{Azuma2002}.
  Let $Y_i$ be an $i^{\U{th}}$ random variable with $Y_0:=0$ 
  ($0\leq i\leq n$) defined on the probability space
  $(\Omega,\mathcal{F},p)$ with the filtration $\mathcal{F}_0\subset\mathcal{F}_1\subset\cdot\cdot\cdot\subset \mathcal{F}_n=\mathcal{F}$, where
  $\mathcal{F}_0$ is the trivial $\sigma$-algebra $\{\emptyset,\Omega\}$ and $\mathcal{F}$ is the power set of $\Omega$ (consisting of all the subsets of $\Omega$). 
  We suppose that the sequence of random variables $\{Y_j\}^n_{j=1}$ satisfies
  \begin{align}
    E[Y_j|\mathcal{F}_{j}]=Y_{j},
      \label{Fjmeasurable}
    \end{align}
  the Martingale condition:
\begin{align}
  E[Y_j|\mathcal{F}_{j-1}]=Y_{j-1},
  \label{martingale}
\end{align}
the bounded difference condition:
\begin{align}
  |Y_j-Y_{j-1}|\leq 1,
  \label{assDelta1}
\end{align}
and the zero-difference condition:
\begin{align}
  p(|Y_j-Y_{j-1}|=0|\mathcal{F}_{j-1})\ge 1-q
  \label{assDelta2}
\end{align}
for a constant $q$ $(0\leq q\leq 1)$.
Then, for any $\epsilon>0$, we have that
\begin{align}
p\Big(Y_n\geq g_{\U{MA}}(\epsilon,q,n)\Big)\leq \epsilon
  \label{newazuma1}
\end{align}
and
\begin{align}
p\Big(Y_n\leq -g_{\U{MA}}(\epsilon,q,n)\Big)\leq \epsilon
  \label{newazuma2}
\end{align}
are satisfied, where $g_{\U{MA}}(\epsilon,q,n)$ is defined by
\begin{align}
  g_{\U{MA}}(\epsilon,q,n):=\frac{\sqrt{\ln\epsilon(\ln\epsilon-18nq)}-\ln\epsilon}{3}. 
  \label{defgA}
\end{align}
\label{th2}
\end{theorem}
For completeness, below we prove Theorem~\ref{LemmaAzumaGeneral}. 
\\\\
{\bf Proof of Theorem~\ref{th2}}~~For convenience, we define $\Delta_j:=Y_j-Y_{j-1}$, and we have that for any $s>0$
\begin{align}
  E[e^{s\Delta_j}|\mathcal{F}_{j-1}]&=E[1+s\Delta_j+s^2\Delta^2_jg(s\Delta_j)|\mathcal{F}_{j-1}]\\
  &= 1+s^2E[\Delta^2_jg(s\Delta_j)|\mathcal{F}_{j-1}]
  \label{martingale2}\\
  &\leq 1+s^2g(s)E[\Delta^2_j|\mathcal{F}_{j-1}]
  \label{Deltaleq1}\\
  &\leq 1+s^2g(s)q
  \label{square}\\
  &\leq e^{s^2g(s)q}\label{result:deltabound},
\end{align}
where $g(x):=(e^x-1-x)/x^2$ with $g(0)=1/2$. 
Here, in Eq.~(\ref{martingale2}) we use Eqs.~(\ref{Fjmeasurable}) and (\ref{martingale}),
in Eq.~(\ref{Deltaleq1}) we use Eq.~(\ref{assDelta1}) and the fact that $g(x)$ is a monotonically increasing function, and finally in Eq.~(\ref{square}) we use Eqs.~(\ref{assDelta1}) and
(\ref{assDelta2}).
We also have that for any $t>0$,
\begin{align}
  p(Y_n\geq t)&\leq e^{-st}E[e^{sY_n}]
  \label{markov}\\
  &=e^{-st}E[E[e^{s(Y_{n-1}+\Delta_n)}|\mathcal{F}_{n-1}]]
  \label{tower}\\
  &=e^{-st}E[e^{sY_{n-1}}E[e^{s\Delta_n}|\mathcal{F}_{n-1}]]
  \label{condition1used}\\
  &\leq e^{s^2g(s)q}e^{-st}E[e^{sY_{n-1}}]
  \label{deltabound}\\
  &\leq\cdot\cdot\cdot\label{cdot}\\
  &\leq e^{-st}\Pi^n_{i=1}e^{s^2g(s)q}\label{Pi}\\
  &= e^{-st+(e^s-1-s)qn}.
  \label{az:final}
\end{align}
Here, Eq.~(\ref{markov}) is due to the Markov's inequality,
Eq.~(\ref{tower}) is due to the tower rule of expectation,
Eq.~(\ref{condition1used}) is due to Eq.~(\ref{Fjmeasurable}), 
Eq.~(\ref{deltabound}) is due to Eq.~(\ref{result:deltabound}),
and in Eq.~(\ref{cdot}) we just employ the same discussion that we used to go from Eq.~(\ref{tower}) to Eq.~(\ref{deltabound}) to
bound $E[e^{sY_{n-1}}]$. 
If we choose $s=\ln(1+\frac{t}{nq})$, we obtain from Eq.~(\ref{az:final}) that
\begin{align}
  p(Y_n\geq t)&\leq e^{-nq\phi(\frac{t}{nq})}
  \label{proof2}
\end{align}
with $\phi(x):=(1+x)\ln(1+x)-x$.
Finally, since $\phi(x)\geq \frac{x^2}{2(1+x/3)}$ for $x>-1$, we obtain
\begin{align}
  p(Y_n\geq t)&\leq \exp\Big(-\frac{t^2}{2nq+2t/3}\Big).
\end{align}
By setting $\exp\Big(-\frac{t^2}{2nq+2t/3}\Big)=\epsilon$, we obtain Eq.~(\ref{newazuma1}).
Also, Eq.~(\ref{newazuma2}) can be proven in a similar way.\sq

\section{Decoy-state method}
\label{sec:appC}
In this Appendix, we present the decoy-state method that we use to estimate bounds on the parameters $S_{c,n,u,\U{det},y,X}$ and 
$S_{Z,Z,n,u,\U{det}}$ for $n=1$ in the presence of intensity fluctuations within the framework of SCIC.
Here, $S_{c,n,u,\U{det},y,X}$ denotes the cardinality of the set $\mathcal{S}_{c,n,u,\U{det},y,X}=\{i|c^{i}=c,n^{i}=n,t^{i}=u,b^{i}=X,y^{i}=y\}$ (with $y\in\{0,1\}$).
More precisely, the goal here is to bound the quantities
\begin{align}
S_{c,n=1,u,\U{det},y,X}~~\U{and}~~S_{Z,Z,n=1,u,\U{det}}
\end{align}
with the available experimentally observed data, that is,
\begin{align}
\{S_{c,k,\U{det},y,X}\}_{k\in\mathcal{K}}~~\U{and}~~\{S_{Z,Z,k,\U{det}}\}_{k\in\mathcal{K}},
\end{align}
respectively. We first derive the upper and lower bounds on $S_{c,1,u,\U{det},y,X}$ (see Sec.~\ref{subsecc}),
and after that we derive the lower bound on $S_{Z,Z,1,u,\U{det}}$ (see Sec.~\ref{subsecZ}).

To achieve this goal, we divide the task into four steps.
To illustrate the method, let us consider for instance the case for $S_{c,1,u,\U{det},y,X}$. 
In step~1, we use the Modified Azuma's inequality to relate
the numbers $S_{c,1,u,\U{det},y,X}$ and $S_{c,k,u,\U{det},y,X}$ with the quantities $\expect{S_{c,1,u,y,X}}_{\U{det}}$ in  Eq.~(\ref{cundet}) and
$\expect{S_{c,k,u,y,X}}_{\U{det}}$ in Eq.~(\ref{cukdetu}), respectively.
Here, $S_{c,k,u,\U{det},y,X}$ denotes the cardinality of the set 
$\mathcal{S}_{c,k,u,\U{det},y,X}=\{i|c^{i}=c,k^{i}=k,t^{i}=u,b^{i}=X,y^{i}=y\}$. 
In step~2, we derive the bounds on $\expect{S_{c,k,u,y,X}}_{\U{det}}$ by using the functions $\{f_{c,n,u,y,X|\U{det}}\}_n$ 
that are related to the quantities $\{\expect{S_{c,n,u,y,X}}_{\U{det}}\}_n$.
Here, $\expect{S_{c,n,u,y,X}}_{\U{det}}$ and $f_{c,n,u,y,X|\U{det}}$ are defined in Eqs.~(\ref{cundet}) and (\ref{deffc}) respectively. 
In step~3, we derive the bounds on $\expect{S_{c,n=1,u,y,X}}_{\U{det}}$ with $\{\expect{S_{c,k,u,y,X}}_{\U{det}}\}_k$. 
In step~4, we finally combine Eqs.~(\ref{S1Upc}) and (\ref{S1Lowc}) with Eqs.~(\ref{Azumackudet}) and (\ref{Azumacu1det})
which are obtained by means of the Modified Azuma's inequality, and 
we obtain the bounds on $S_{c,n=1,u,\U{det},y,X}$ with the observed data $\{S_{c,k,\U{det},y,X}\}_{k\in\mathcal{K}}$.

\subsection{Upper and lower bounds on $S_{c,1,u,\U{det},y,X}$}
\label{subsecc}
\subsubsection{Step 1. Modified Azuma's inequality}
\label{sec:appC1}
In order to obtain upper and lower bounds on $S_{c,1,u,\U{det},y,X}$, 
we first relate the numbers $S_{c,1,u,\U{det},y,X}$ and $S_{c,k,u,\U{det},y,X}$ with the quantities
$\expect{S_{c,1,u,y,X}}_{\U{det}}$ and $\expect{S_{c,k,u,y,X}}_{\U{det}}$, respectively. 
In the discussion, we use a shorthand notation
$\bm{\chi}^{i-1}_{AB}:=\bm{\chi}_A^{i-1},\bm{\chi}_B^{i-1}$ with $\bm{\chi}_A^{i-1}=(\bm{c}^{i-1},\bm{k}^{i-1},\bm{g}^{i-1},\bm{n}^{i-1})$
and $\bm{\chi}_B^{i-1}=(\bm{b}^{i-1},\bm{y}^{i-1})$, and $\expect{S_{c,k,u,y,X}}_{\U{det}}$ is defined as
\begin{align}
  \expect{S_{c,k,u,y,X}}_{\U{det}}
  =\sum_{i\in\mathcal{S}_{\U{det}}}p(c^{i}=c,k^{i}=k,t^{i}=u,y^{i}=y|\bm{\chi}^{i-1}_{AB},y^{i}\neq\emptyset,b^{i}=X)\delta(b^{i},X).
  \label{cukdetu}
\end{align}
Here, $\delta(x,y)$ denotes the Kronecker delta. 
To relate Eq.~(\ref{cukdetu}) with the number $S_{c,k,u,\U{det},y,X}$, we introduce the following random variable
\begin{align}
  Y^{j}_{c,k,u,y,X}:=\Lambda^{j}_{c,k,u,y,X}-
    \sum_{i\in\mathcal{S}^{j}_{\U{det}}}
    p(c^{i}=c,k^{i}=k,t^{i}=u,y^{i}=y|\bm{\chi}^{i-1}_{AB},y^{i}\neq\emptyset,b^{i}=X)\delta(b^{i},X),
  \label{Yc}
\end{align}
where $\mathcal{S}^{j}_{\U{det}}$ is the set containing the first $j$ elements of the set $\mathcal{S}_{\U{det}}$, and 
$\Lambda^{j}_{c,k,u,y,X}$ denotes the number of instances where 
$(c,k,u,X,y)$ is realised among the set $\mathcal{S}^{j}_{\U{det}}$, namely,
$\Lambda^{j}_{c,k,u,y,X}=\sum_{i\in\mathcal{S}^{j}_{\U{det}}}\delta(c^i,c)\delta(k^i,k)\delta(t^i,u)\delta(y^i,y)\delta(b^i,X)$. 
Then, the random variables $\{Y^{j}_{c,k,u,y,X}\}^{N_{\U{det}}}_{j=1}$ satisfy Eq.~(\ref{Fjmeasurable}),
the Martingale condition [see Eq.~(\ref{martingale})],
the bounded difference condition [see Eq.~(\ref{assDelta1})] and the zero-difference condition [see Eq.~(\ref{assDelta2})]
\footnote{
  This statement can be proven as follows. Since $Y^{j}_{c,k,u,y,X}-Y^{j-1}_{c,k,u,y,X}$ is written as
  \begin{align}
    Y^{j}_{c,k,u,y,X}-Y^{j-1}_{c,k,u,y,X}=\sum_{i\in \mathcal{S}^{j}_{\U{det}}\setminus \mathcal{S}^{j-1}_{\U{det}}}
    \left[\delta(c^i,c)\delta(k^i,k)\delta(t^i,u)\delta(y^i,y)-p(c^{i}=c,k^{i}=k,t^{i}=u,y^{i}=y|\bm{\chi}^{i-1}_{AB},y^{i}\neq\emptyset,b^i=X)
      \right]\delta(b^i,X),
    \label{differenceJJ-1}
  \end{align}
  we can confirm that the Martingale condition
\begin{align}
      E[Y_j-Y_{j-1}|\mathcal{F}_{j-1}]&=E\left[\sum_{i\in \mathcal{S}^{j}_{\U{det}}\setminus \mathcal{S}^{j-1}_{\U{det}}}
    \left[\delta(c^i,c)\delta(k^i,k)\delta(t^i,u)\delta(y^i,y)-p(c^{i}=c,k^{i}=k,t^{i}=u,y^{i}=y|\bm{\chi}^{i-1}_{AB},y^{i}\neq\emptyset,b^i=X)
      \right]\delta(b^i,X)\middle|\mathcal{F}_{j-1}\right]\notag\\
      &=\sum_{i\in \mathcal{S}^{j}_{\U{det}}\setminus \mathcal{S}^{j-1}_{\U{det}}}
      p(c^{i}=c,k^{i}=k,t^{i}=u,y^{i}=y|\bm{\chi}^{i-1}_{AB},y^{i}\neq\emptyset,b^i=X)
      p(b^i=X|\bm{\chi}^{i-1}_{AB},y^{i}\neq\emptyset)\notag\\
      &-\sum_{i\in \mathcal{S}^{j}_{\U{det}}\setminus \mathcal{S}^{j-1}_{\U{det}}}
      p(c^{i}=c,k^{i}=k,t^{i}=u,y^{i}=y|\bm{\chi}^{i-1}_{AB},y^{i}\neq\emptyset,b^i=X)p(b^i=X|\bm{\chi}^{i-1}_{AB},y^{i}\neq\emptyset)
      \notag\\
      &=0\notag
  \end{align}
  holds. 
  Also, Eq.~(\ref{differenceJJ-1}) assures the bounded difference condition: $|Y_j-Y_{j-1}|\le1$.
  Finally, from Eq.~(\ref{differenceJJ-1}) and using the assumptions (B-1) and (B-2), we obtain the zero-difference condition in Eq.~(\ref{zerod}) as
  $p(|Y^{j}_{c,k,u,y,X}-Y^{j-1}_{c,k,u,y,X}|=0|\mathcal{F}_{j-1})\ge p(\sum_{i\in \mathcal{S}^{j}_{\U{det}}\setminus \mathcal{S}^{j-1}_{\U{det}}}\delta(b^i,X)=0|\mathcal{F}_{j-1})=1-p^B_X$.
}
:
\begin{align}
  p(|Y^{j}_{c,k,u,y,X}-Y^{j-1}_{c,k,u,y,X}|=0|\mathcal{F}_{j-1})\ge 1-p^{B}_X.
  \label{zerod}
\end{align}
Here, we identify the filtration $\mathcal{F}_{j-1}$ with the random variables $(\bm{\chi}^{\max\{i|i\in\mathcal{S}^j_{\U{det}}\}-1}_{AB})$. 
Therefore, we can apply the Modified Azuma's inequality in Eqs.~(\ref{newazuma1}) and (\ref{newazuma2}), and we obtain
\begin{align}
  p\Big(S_{c,k,u,\U{det},y,X}-\expect{S_{c,k,u,y,X}}_{\U{det}}\geq
  g_{\U{MA}}(\epsilon^{c,k,u,y,X}_{\U{MA}},p^{B}_X,N_{\U{det}})\Big)\leq \epsilon^{c,k,u,y,X}_{\U{MA}}
  \label{MA1}
\end{align}
and
\begin{align}
  p\Big(S_{c,k,u,\U{det},y,X}-\expect{S_{c,k,u,y,X}}_{\U{det}}\leq -g_{\U{MA}}(\epsilon^{c,k,u,y,X}_{\U{MA}},p^{B}_X,N_{\U{det}})\Big)\leq \epsilon^{c,k,u,y,X}_{\U{MA}}.
    \label{MA2}
\end{align}
Here, $\epsilon^{c,k,u,y,X}_{\U{MA}}$ denotes the failure probability associated to the estimation of $\expect{S_{c,k,u,y,X}}_{\U{det}}$, 
and the function $g_{\U{MA}}(\epsilon^{c,k,u,y,X}_{\U{MA}},p^{B}_X,N_{\U{det}})$ is defined in Eq.~(\ref{defgA}). 
Note that the quantity $S_{c,k,u,\U{det},y,X}$ in Eqs.~(\ref{MA1}) and (\ref{MA2}) is not directly observed in the actual experiments,
and we only know its range, given that the number of tagged signals $n_{\U{tag}}$ is upper bounded by $N_{\U{tag}}$, as
\begin{align}
  S^-_{c,k,u,\U{det},y,X}:=
  S_{c,k,\U{det},y,X}-N_{\U{tag}}\leq S_{c,k,u,\U{det},y,X}\leq S_{c,k,\U{det},y,X}, 
  \label{sun_mc}
\end{align}
where $S_{c,k,\U{det},y,X}:=|\mathcal{S}_{c,k,\U{det},y,X}|$.
By combining Eqs.~(\ref{MA1}), (\ref{MA2}) and (\ref{sun_mc}), we obtain
\begin{align}
S^-_{c,k,u,\U{det},y,X}-g_{\U{MA}}(\epsilon^{c,k,u,y,X}_{\U{MA}},p^{B}_X,N_{\U{det}})
\leq \expect{S_{c,k,u,y,X}}_{\U{det}} \leq S_{c,k,\U{det},y,X}+g_{\U{MA}}(\epsilon^{c,k,u,y,X}_{\U{MA}},p^{B}_X,N_{\U{det}})
\label{Azumackudet}
\end{align}
except for error probability $2\epsilon^{c,k,u,y,X}_{\U{MA}}$.
The bounds on $S_{c,n=1,u,\U{det},y,X}$ will be expressed as a function of the rhs and the lhs of Eq.~(\ref{Azumackudet}).

Next, we define the quantity $\expect{S_{c,n,u,y,X}}_{\U{det}}$ that will be related to the number $S_{c,n,u,\U{det},y,X}$ through the
Modified Azuma's inequality as
\begin{align}
  \expect{S_{c,n,u,y,X}}_{\U{det}}:=\sum_{i\in\mathcal{S}_{\U{det}}}p(c^{i}=c,n^{i}=n,t^{i}=u,y^{i}=y|\bm{\chi}^{i-1}_{AB},y^{i}\neq\emptyset,b^i=X)
  \delta(b^{i},X),
  \label{cundet}
\end{align}
and we follow the above arguments about the Modified Azuma's inequality. As a result, we obtain 
\begin{align}
  \expect{S_{c,n,u,y,X}}_{\U{det}}-g_{\U{MA}}(\epsilon^{c,n,u,y,X}_{\U{MA}},p^{B}_X,N_{\U{det}})
\le S_{c,n,u,\U{det},y,X}\le\expect{S_{c,n,u,y,X}}_{\U{det}}+g_{\U{MA}}(\epsilon^{c,n,u,y,X}_{\U{MA}},p^{B}_X,N_{\U{det}})
\label{Azumacu1det}
\end{align}
except for error probability $2\epsilon^{c,n,u,y,X}_{\U{MA}}$. 

From Eqs.~(\ref{Azumackudet}) and (\ref{Azumacu1det}) obtained by means of the Modified Azuma's inequality, we find that 
the remaining task is to bound $\expect{S_{c,1,u,y,X}}_{\U{det}}$ with $\{\expect{S_{c,k,u,y,X}}_{\U{det}}\}_k$ such that
$S_{c,1,u,\U{det},y,X}$ can be bounded by the observed data appearing in Eq.~(\ref{Azumackudet}).
Next, in step~2, we relate the expectation $\expect{S_{c,k,u,y,X}}_{\U{det}}$ with $\{\expect{S_{c,n,u,y,X}}_{\U{det}}\}_n$
through some functions $\{f_{c,n,u,y,X|\U{det}}\}_n$, 
and then by using this relation we bound $\expect{S_{c,1,u,y,X}}_{\U{det}}$ with $\{\expect{S_{c,k,u,y,X}}_{\U{det}}\}_k$ in step~3.

\subsubsection{Step 2. Upper and lower bounds on $\expect{S_{c,k,u,y,X}}_{\U{det}}$ with $\{f_{c,n,u,y,X|\U{det}}\}_n$}
\label{sec:appC2}
In this step, we bound $\expect{S_{c,k,u,y,X}}_{\U{det}}$ in Eq.~(\ref{cukdetu}) with $\{\expect{S_{c,n,u,y,X}}_{\U{det}}\}_n$
through the functions $\{f_{c,n,u,y,X|\U{det}}\}_n$. 
In the following discussion, we use the notation $\bm{g}^{\ge i}:=g^{i},g^{i+1},...,g^{N_{\U{sent}}}$. 

First, from the law of total probability, the definition (D-1) in Sec.~\ref{Assumptions on the devices} and Bayes' theorem,
Eq.~(\ref{cukdetu}) can be rewritten as
\begin{align}
  \expect{S_{c,k,u,y,X}}_{\U{det}}=
  &\sum_{i\in\mathcal{S}_{\U{det}}}\sum_n\sum_{g^i\in\mathcal{G}^i_{\U{unt}}}\sum_{\bm{g}^{\ge i+1}}
  p(c^{i}=c,k^{i}=k,n^{i}=n,\bm{g}^{\ge i},y^{i}=y|\bm{\chi}^{i-1}_{AB},y^{i}\neq\emptyset,b^i=X)\delta(b^i,X)\notag\\
    =&\sum_{i\in\mathcal{S}_{\U{det}}}\sum_n\sum_{g^i\in\mathcal{G}^i_{\U{unt}}}\sum_{\bm{g}^{\ge i+1}}
  p(k^{i}=k|c^{i}=c,n^{i}=n,\bm{g}^{\ge i},y^i\neq \emptyset,y^{i}=y,\bm{\chi}^{i-1}_{AB},b^i=X)\notag\\
  &\times p(c^{i}=c,n^{i}=n,\bm{g}^{\ge i},y^{i}=y|\bm{\chi}^{i-1}_{AB},y^{i}\neq\emptyset,b^i=X)\delta(b^i,X).
  \label{1c}
\end{align}
Next, we calculate the first factor of Eq.~(\ref{1c}) to obtain Eq.~(\ref{apC:sixth}) below. 
\begin{align}
  &p(k^{i}=k|c^{i}=c,n^{i}=n,\bm{g}^{\ge i},y^i\neq \emptyset,y^{i}=y,\bm{\chi}^{i-1}_{AB},b^i=X)\notag\\
  =&p(k^{i}=k|c^{i}=c,n^{i}=n,\bm{g}^{\ge i},\bm{\chi}^{i-1}_{AB},b^i=X)\label{apC:first}\\
=&p(k^{i}=k|c^{i}=c,n^{i}=n,\bm{g}^{\ge i},\bm{\chi}_{AB}^{i-1})\label{apC:second}\\
  =&p(k^{i}=k|c^{i}=c,n^{i}=n,\bm{g}^{\ge i},\bm{\chi}_{A}^{i-1})\label{apC:second2}\\
=&\frac{p(n^{i}=n|c^{i}=c,k^{i}=k,\bm{g}^{\ge i},\bm{\chi}_{A}^{i-1})
  p(k^{i}=k|c^{i}=c,\bm{g}^{\ge i},\bm{\chi}_A^{i-1})}
  {p(n^{i}=n|c^{i}=c,\bm{g}^{\ge i},\bm{\chi}_A^{i-1})}\label{apC:third}\\
  =&\frac{p_kp(n^{i}=n|c^{i}=c,k^{i}=k,\bm{g}^{\ge i},\bm{\chi}_A^{i-1})}
  {\sum_{k'}p_{k'}p(n^{i}=n|c^{i}=c,k^{i}=k',\bm{g}^{\ge i},\bm{\chi}_A^{i-1})}\label{apC:fourth}\\
  =&\frac{p_k\sum_{\mu^i}p(n^{i}=n|c^{i}=c,k^{i}=k,\bm{g}^{\ge i},\bm{\chi}_A^{i-1},\mu^{i})
    p(\mu^{i}|c^{i}=c,k^{i}=k,\bm{g}^{\ge i},\bm{\chi}_A^{i-1})}
  {\sum_{k'}p_{k'}\sum_{\mu^i}p(n^{i}=n|c^{i}=c,k^{i}=k',\bm{g}^{\ge i},\bm{\chi}_A^{i-1},\mu^{i})
    p(\mu^{i}|c^{i}=c,k^{i}=k',\bm{g}^{\ge i},\bm{\chi}_A^{i-1})}\label{apC:fifth}\\
  =&\frac{p_k\sum_{\mu^i}p(n^{i}=n|\mu^{i})\delta(\mu^{i},\mu^{i}_{k,g^i})}
  {\sum_{k'}p_{k'}\sum_{\mu^i}p(n^{i}=n|\mu^{i})\delta(\mu^{i},\mu^{i}_{k',g^i})},
    \label{apC:sixth}
\end{align}
where in Eq.~(\ref{apC:first}) we use the decoy-state property, namely,
$p(y^i\neq \emptyset,y^{i}=y|k^{i}=k,c^{i}=c,n^{i}=n,\bm{g}^{\geq i},\bm{\chi}^{i-1}_{AB},b^i=X)
=p(y^i\neq \emptyset,y^{i}=y|c^{i}=c,n^{i}=n,\bm{g}^{\geq i},\bm{\chi}^{i-1}_{AB},b^i=X)$ holds.
This is because once $\bm{g}^{{N_{\U{sent}}}}$ is fixed, Eq.~(\ref{sendingrho}) assures that the total $N_{\U{sent}}$ sending states
have a tensor product structure which implies that the information of $k^i$ is only encoded in the $i^{\U{th}}$ sending system $B^i$, 
and the $n$-photon state in the system $B^i$ is the same for any choice of $k^{i}$. 
In Eq.~(\ref{apC:second}) we use the assumption (B-2) in Sec.~\ref{assbob},
in Eq.~(\ref{apC:second2}) we recursively use the decoy-state property and the assumption (B-2), 
in Eqs.~(\ref{apC:third}) and (\ref{apC:fifth}) we use the Bayes' theorem, and finally in Eqs.~(\ref{apC:fourth}) and (\ref{apC:sixth}) we use Eq.~(\ref{sendingrho}). 

In Eq.~(\ref{apC:sixth}), from Eq.~(\ref{Rint}), $\forall i\in \mathcal{S}_{\U{det}}$ and $\forall g^i\in\mathcal{G}^i_{\U{unt}}$, 
$\sum_{\mu^i}p(n^i=n|\mu^i)\delta(\mu^{i},\mu^{i}_{k,g^i})$ can be upper [lower] bounded by
$e^{-\mu^-_k}$ $[e^{-\mu^+_k}]$ for $n=0$ and $\frac{e^{-\mu^+_k}(\mu^+_k)^n}{n!}$ [$\frac{e^{-\mu^-_k}(\mu^-_k)^n}{n!}$] for
$n\ge 1$, which we denote by ${\rm Pois}^{\U{U}}(n|k,u)$ $[{\rm Pois}^{\U{L}}(n|k,u)]$.
With these bounds, $\expect{S_{c,k,u,y,X}}_{\U{det}}$ can be upper and lower bounded as 
\begin{align}
  \expect{S_{c,k,u,y,X}}_{\U{det}}\leq&p_k
  \sum_n{\rm Pois}^{\U{U}}(n|k,u)\sum_{i\in\mathcal{S}_{\U{det}}}\sum_{g^i\in\mathcal{G}^i_{\U{unt}}}
  \frac{p(c^{i}=c,n^{i}=n,g^{i},y^{i}=y|\bm{\chi}^{i-1}_{AB},y^{i}\neq\emptyset,b^i=X)}
       {\sum_{k'}p_{k'}\sum_{\mu^i}p(n^{i}=n|\mu^{i})\delta(\mu^{i},\mu^{i}_{k',g^i})}\delta(b^i,X)\notag\\
  =&p_k\sum_n{\rm Pois}^{\U{U}}(n|k,u)f_{c,n,u,y,X|\U{det}},
  \label{4c}
\end{align}
where we define the function
\begin{align}
f_{c,n,u,y,X|\U{det}}=\sum_{i\in\mathcal{S}_{\U{det}}}\sum_{g^i\in\mathcal{G}^i_{\U{unt}}}
\frac{p(c^{i}=c,n^{i}=n,g^{i},y^{i}=y|\bm{\chi}^{i-1}_{AB},y^{i}\neq\emptyset,b^i=X)}
     {\sum_{k'}p_{k'}\sum_{\mu^i}p(n^{i}=n|\mu^{i})\delta(\mu^{i},\mu^{i}_{k',g^i})}\delta(b^i,X)
  \label{deffc}
  \end{align}
and
\begin{align}
  \expect{S_{c,k,u,y,X}}_{\U{det}}\geq p_k\sum_n{\rm Pois}^{\U{L}}(n|k,u)f_{c,n,u,y,X|\U{det}},
  \label{5c}
\end{align}
respectively.
It is notable that Eq.~(\ref{4c}) represents a linear combination of $\{f_{c,n,u,y,X|\U{det}}\}_n$, and
$f_{c,n,u,y,X|\U{det}}$ is {\it independent} of $k$, which is crucial in the next discussion. 

\subsubsection{Step 3. Upper and lower bounds on $\expect{S_{c,n=1,u,y,X}}_{\U{det}}$ with $\{\expect{S_{c,k,u,y,X}}_{\U{det}}\}_k$}
\label{sec:appC3}
In this step, we derive an upper and lower bounds on $\expect{S_{c,n=1,u,y,X}}_{\U{det}}$ with $\{\expect{S_{c,k,u,y,X}}_{\U{det}}\}_k$. 

\textbf{(i) Upper bound on} $\bm{\expect{S_{c,1,u,y,X}}_{\U{det}}}$

First, we derive an upper bound on $\expect{S_{c,1,u,y,X}}_{\U{det}}$. 
From Eqs.~(\ref{4c}) and (\ref{5c}), we have two inequalities:
$\expect{S_{c,k_2,u,y,X}}_{\U{det}}\geq p_{k_2}\sum_n{\rm Pois}^{\U{L}}(n|k_2,u)f_{c,n,u,y,X|\U{det}}$ and
$\expect{S_{c,k_3,u,y,X}}_{\U{det}}\leq p_{k_3}\sum_n{\rm Pois}^{\U{U}}(n|k_3,u)f_{c,n,u,y,X|\U{det}}$.
By combining them  and using the conditions $1>\mu^+_{k_2}$ and $\mu^-_{k_2}>\mu^+_{k_3}$ (which we actually imposed in the protocol), 
$f_{c,1,u,y,X|\U{det}}$ is upper bounded by
\begin{align}
  f_{c,1,u,y,X|\U{det}}\leq\frac{\expect{S_{c,k_2,u,y,X}}_{\U{det}}{\rm Pois}^{\U{U}}(0|k_3,u)/p_{k_2}
    -\expect{S_{c,k_3,u,y,X}}_{\U{det}}{\rm Pois}^{\U{L}}(0|k_2,u)/p_{k_3}}
  {{\rm Pois}^{\U{L}}(1|k_2,u){\rm Pois}^{\U{U}}(0|k_3,u)-{\rm Pois}^{\U{U}}(1|k_3,u){\rm Pois}^{\U{L}}(0|k_2,u)}.
\end{align}
Substituting the definition in Eq.~(\ref{deffc}) gives the upper bound on $\expect{S_{c,1,u,y,X}}_{\U{det}}$ as
\begin{align}
  \expect{S_{c,1,u,y,X}}_{\U{det}}=&
  \sum_{i\in\mathcal{S}_{\U{det}}}p(c^{i}=c,n^{i}=1,t^{i}=u,y^{i}=y|\bm{\chi}_{AB}^{i-1},y^{i}\neq\emptyset,b^i=X)\delta(b^{i},X)\notag\\
  \leq&\frac{\expect{S_{c,k_2,u,y,X}}_{\U{det}}{\rm Pois}^{\U{U}}(0|k_3,u)/p_{k_2}-
    \expect{S_{c,k_3,u,y,X}}_{\U{det}}{\rm Pois}^{\U{L}}(0|k_2,u)/p_{k_3}}
      {{\rm Pois}^{\U{L}}(1|k_2,u){\rm Pois}^{\U{U}}(0|k_3,u)-{\rm Pois}^{\U{U}}(1|k_3,u){\rm Pois}^{\U{L}}(0|k_2,u)}{\rm Pois}^{\U{U}}(1|u).
      \label{S1Upc}
\end{align}
Here, we define ${\rm Pois}^{\U{U}}(1|u)=\sum_{k\in\mathcal{K}}p_k{\rm Pois}^{\U{U}}(1|k,u)$.

\textbf{(ii) Lower bound on} $\bm{\expect{S_{c,1,u,y,X}}_{\U{det}}}$

Next, we derive a lower bound on $\expect{S_{c,1,u,y,X}}_{\U{det}}$. 
For this, we first use Eqs.~(\ref{4c}) and (\ref{5c}), and we have three inequalities: 
\begin{align}
  &e^{\mu^-_{k_2}}\expect{S_{c,k_2,u,y,X}}_{\U{det}}\leq e^{\mu^-_{k_2}}p_{k_2}\sum_n{\rm Pois}^{\U{U}}(n|k_2,u)f_{c,n,u,y,X|\U{det}},\notag\\
  &e^{\mu^+_{k_3}}\expect{S_{c,k_3,u,y,X}}_{\U{det}}\geq e^{\mu^+_{k_3}}p_{k_3}\sum_n{\rm Pois}^{\U{L}}(n|k_3,u)f_{c,n,u,y,X|\U{det}},\notag\\
  &e^{\mu^+_{k_1}}\expect{S_{c,k_1,u,y,X}}_{\U{det}}\geq e^{\mu^+_{k_1}}p_{k_1}\sum_n{\rm Pois}^{\U{L}}(n|k_1,u)f_{c,n,u,y,X|\U{det}}.
  \label{ineqiiic}
\end{align}
By combining the first two inequalities, it is easy to show that
\begin{align}
  (\mu^+_{k_2}-\mu^-_{k_3})f_{c,1,u,y,X|\U{det}}
  \geq \frac{e^{\mu^-_{k_2}}
    \expect{S_{c,k_2,u,y,X}}_{\U{det}}}{p_{k_2}}
  -\frac{e^{\mu^+_{k_3}}\expect{S_{c,k_3,u,y,X}}_{\U{det}}}{p_{k_3}}-\sum^{\infty}_{n=2}\frac{(\mu^+_{k_2})^n-(\mu^-_{k_3})^n}{n!}
  f_{c,n,u,y,X|\U{det}}.
  \label{easyc}
\end{align}
If we assume $\mu^-_{k_1}>\mu^+_{k_2}+\mu^+_{k_3}$ (which we actually imposed in the protocol), we have that 
$(\mu_{k_1}^-)^2[({\mu_{k_2}^+})^n-({\mu_{k_3}^-})^n]\leq[(\mu_{k_2}^+)^2-(\mu_{k_3}^-)^2](\mu_{k_1}^-)^{n}$ holds for
$n\geq2$~\cite{PhysRevA.72.012326}. Therefore, Eq.~(\ref{easyc}) leads to
\begin{align}
  (\mu^+_{k_2}-\mu^-_{k_3})f_{c,1,u,y,X|\U{det}}\geq
  \frac{e^{\mu^-_{k_2}}\expect{S_{c,k_2,u,y,X}}_{\U{det}}}{p_{k_2}}
  -\frac{e^{\mu^+_{k_3}}\expect{S_{c,k_3,u,y,X}}_{\U{det}}}{p_{k_3}}-\frac{(\mu^+_{k_2})^2-(\mu^-_{k_3})^2}{(\mu^-_{k_1})^2}
  \sum^{\infty}_{n=2}\frac{(\mu^-_{k_1})^n}{n!}f_{c,n,u,y,X|\U{det}},
\end{align}
which is equivalent to
\begin{align}
  f_{c,1,u,y,X|\U{det}}&\geq \frac{\mu^-_{k_1}}{(\mu^+_{k_2}-\mu^-_{k_3})(\mu^-_{k_1}-\mu^+_{k_2}-\mu^-_{k_3})}
  \Big[\frac{e^{\mu^-_{k_2}}\expect{S_{c,k_2,u,y,X}}_{\U{det}}}{p_{k_2}}
  -\frac{e^{\mu^+_{k_3}}\expect{S_{c,k_3,u,y,X}}_{\U{det}}}{p_{k_3}}\notag\\
  &-\frac{(\mu^+_{k_2})^2-(\mu^-_{k_3})^2}{(\mu^-_{k_1})^2}\Big(
  \sum^{\infty}_{n=0}\frac{(\mu^-_{k_1})^n}{n!}f_{c,n,u,y,X|\U{det}}-f_{c,0,u,y,X|\U{det}}\Big)\Big].
  \label{s1c}
\end{align}
If we assume $1>\mu^+_{k_1}$ (which we actually imposed in the protocol), by using in Eq.~(\ref{s1c}), 
the third inequality in Eq.~(\ref{ineqiiic}) and the definition of $f_{c,1,u,y,X|\U{det}}$ in Eq.~(\ref{deffc}), 
we obtain the lower bound on $\expect{S_{c,1,u,y,X}}_{\U{det}}$ as a function of $\{\expect{S_{c,k,u,y,X}}_{\U{det}}\}_k$ as 
\begin{align}
\expect{S_{c,1,u,y,X}}_{\U{det}}
\ge&
\frac{\mu^-_{k_1}{\rm Pois}^{\U{L}}(1|u)}{(\mu^+_{k_2}-\mu^-_{k_3})(\mu^-_{k_1}-\mu^+_{k_2}-\mu^-_{k_3})}
\Big\{
  \frac{e^{\mu^-_{k_2}}\expect{S_{c,k_2,u,y,X}}_{\U{det}}}{p_{k_2}}
  -\frac{e^{\mu^+_{k_3}}\expect{S_{c,k_3,u,y,X}}_{\U{det}}}{p_{k_3}}\notag\\
&-\frac{(\mu^+_{k_2})^2-(\mu^-_{k_3})^2}{(\mu^-_{k_1})^2}
  \Big(
 \frac{e^{\mu^+_{k_1}}\expect{S_{c,k_1,u,y,X}}_{\U{det}}}{p_{k_1}} \Big)\Big\}.
 \label{S1Lowc}
\end{align}
Here, we define ${\rm Pois}^{\U{L}}(1|u)=\sum_{k\in\mathcal{K}}p_k{\rm Pois}^{\U{L}}(1|k,u)$.

\subsubsection{Step 4. Upper and lower bounds on $S_{c,n=1,u,\U{det},y,X}$ with $\{S_{c,k,\U{det},y,X}\}_k$}
\label{sec:appC4c}
In this step, we finally derive the upper and the lower bounds on $S_{c,n=1,u,\U{det},y,X}$ with experimentally observed data $\{S_{c,k,\U{det},y,X}\}_k$.

\textbf{(i) Upper bound on} $\bm{S_{c,1,u,\U{det},y,X}}$

We combine Eqs.~(\ref{Azumackudet}) and (\ref{Azumacu1det}), which result from the application of the Modified Azuma's inequality, with
Eq.~(\ref{S1Upc}) to derive the upper bound on $S_{c,1,u,\U{det},y,X}$. 
\begin{align}
  &S_{c,1,u,\U{det},y,X}\leq S^{\U{U}}_{c,1,u,\U{det},y,X}:=\notag\\
  &\frac{\frac{[S_{c,k_2,\U{det},y,X}+g_{\U{MA}}(\epsilon^{c,k_2,u,y,X}_{\U{MA}},p^{B}_X,N_{\U{det}})]{\rm Pois}^{\U{U}}(0|k_3,u)}{p_{k_2}}-
    \frac{[S^-_{c,k_3,u,\U{det},y,X}-g_{\U{MA}}(\epsilon^{c,k_3,u,y,X}_{\U{MA}},p^{B}_X,N_{\U{det}})]{\rm Pois}^{\U{L}}(0|k_2,u)}{p_{k_3}}}
       {{\rm Pois}^{\U{L}}(1|k_2,u){\rm Pois}^{\U{U}}(0|k_3,u)-{\rm Pois}^{\U{U}}(1|k_3,u){\rm Pois}^{\U{L}}(0|k_2,u)}\notag\\
       &\times {\rm Pois}^{\U{U}}(1|u)+g_{\U{MA}}(\epsilon^{c,1,u,y,X}_{\U{MA}},p^{B}_X,N_{\U{det}})
  \label{eq:resultc1U}
\end{align}
except for error probability
\begin{align}
  \epsilon^{c,1,u,y,X}_{\U{U}}:=\sum_{k=k_2,k_3}\epsilon^{c,k,u,y,X}_{\U{MA}}+\epsilon^{c,1,u,y,X}_{\U{MA}}.
\label{epc1uyxU}  
  \end{align}

\textbf{(ii) Lower bound on} $\bm{S_{c,1,u,\U{det},y,X}}$

Similarly, we combine Eqs.~(\ref{Azumackudet}) and (\ref{Azumacu1det}) with
Eq.~(\ref{S1Lowc}) to derive the lower bound on $S_{c,1,u,\U{det},y,X}$. 
\begin{align}
  &S_{c,1,u,\U{det},y,X}\ge S^{\U{L}}_{c,1,u,\U{det},y,X}:=
\frac{\mu^-_{k_1}{\rm Pois}^{\U{L}}(1|u)}{(\mu^+_{k_2}-\mu^-_{k_3})(\mu^-_{k_1}-\mu^+_{k_2}-\mu^-_{k_3})}
\Big\{
  \frac{e^{\mu^-_{k_2}}[S^-_{c,k_2,u,\U{det},y,X}-g_{\U{MA}}(\epsilon^{c,k_2,u,y,X}_{\U{MA}},p^{B}_X,N_{\U{det}})]}{p_{k_2}}\notag\\
 & -\frac{e^{\mu^+_{k_3}}[S_{c,k_3,\U{det},y,X}+g_{\U{MA}}(\epsilon^{c,k_3,u,y,X}_{\U{MA}},p^{B}_X,N_{\U{det}})]}{p_{k_3}}\notag\\
&-\frac{(\mu^+_{k_2})^2-(\mu^-_{k_3})^2}{(\mu^-_{k_1})^2}
  \Big(\frac{e^{\mu^+_{k_1}}[S_{c,k_1,\U{det},y,X}+g_{\U{MA}}(\epsilon^{c,k_1,u,y,X}_{\U{MA}},p^{B}_X,N_{\U{det}})]}{p_{k_1}}
                \Big)\Big\}+g_{\U{MA}}(\epsilon^{c,1,u,y,X}_{\U{MA}},p^{B}_X,N_{\U{det}})
  \label{eq:resultc1L}
\end{align}
except for error probability
\begin{align}
  \epsilon^{c,1,u,y,X}_{\U{L}}:=\sum_{k\in\mathcal{K}}\epsilon^{c,k,u,y,X}_{\U{MA}}+\epsilon^{c,1,u,y,X}_{\U{MA}}.
  \label{epc1uyxL}  
\end{align}

\subsection{Lower bound on $S_{Z,Z,1,u,\U{det}}$}
\label{subsecZ}
Just by following the same arguments that we use for the estimation of the bounds on $S_{c,1,u,\U{det},y,X}$ in Secs.~\ref{sec:appC1}-
\ref{sec:appC4c}, we find that $S_{Z,Z,1,u,\U{det}}$ is lower bounded by the following quantity:
\begin{align}
  &S^{\U{L}}_{Z,Z,1,u,\U{det}}\notag\\
  &=\frac{\mu^-_{k_1}\sum_{k\in\mathcal{K}}p_k\mu^-_ke^{-\mu^-_k}}{(\mu^+_{k_2}-\mu^-_{k_3})(\mu^-_{k_1}-\mu^+_{k_2}-\mu^-_{k_3})}
\Big\{
  \frac{e^{\mu^-_{k_2}}[S^-_{Z,Z,k_2,u,\U{det}}-g_{\U{MA}}(\epsilon^{Z,k_2,u}_{\U{MA}},p^{B}_Z,N_{\U{det}})]}{p_{k_2}}
-\frac{e^{\mu^+_{k_3}}[S_{Z,Z,k_3,\U{det}}+g_{\U{MA}}(\epsilon^{Z,k_3,u}_{\U{MA}},p^{B}_Z,N_{\U{det}})]}{p_{k_3}}\notag\\
  &-\frac{(\mu^+_{k_2})^2-(\mu^-_{k_3})^2}{(\mu^-_{k_1})^2}\Big(\frac{e^{\mu^+_{k_1}}[S_{Z,Z,k_1,\U{det}}+g_{\U{MA}}(\epsilon^{Z,k_1,u}_{\U{MA}},p^{B}_Z,N_{\U{det}})]}{p_{k_1}}
                \Big)\Big\}+g_{\U{MA}}(\epsilon^{Z,1,u}_{\U{MA}},p^{B}_Z,N_{\U{det}})
  \label{eq:resultZ1L}
\end{align}
except for error probability
\begin{align}
  \epsilon_Z:=\sum_{k\in\mathcal{K}}\epsilon^{Z,k,u}_{\U{MA}}+\epsilon^{Z,1,u}_{\U{MA}}+p_{\U{fail}}.
  \end{align}
Here, we define $S^-_{Z,Z,k_2,u,\U{det}}:=S_{Z,Z,k_2,\U{det}}-N_{\U{tag}}$.

\section{Derivation of the phase error rate}
\label{sec:appD}
\begin{figure}[t]
\includegraphics[width=13cm]{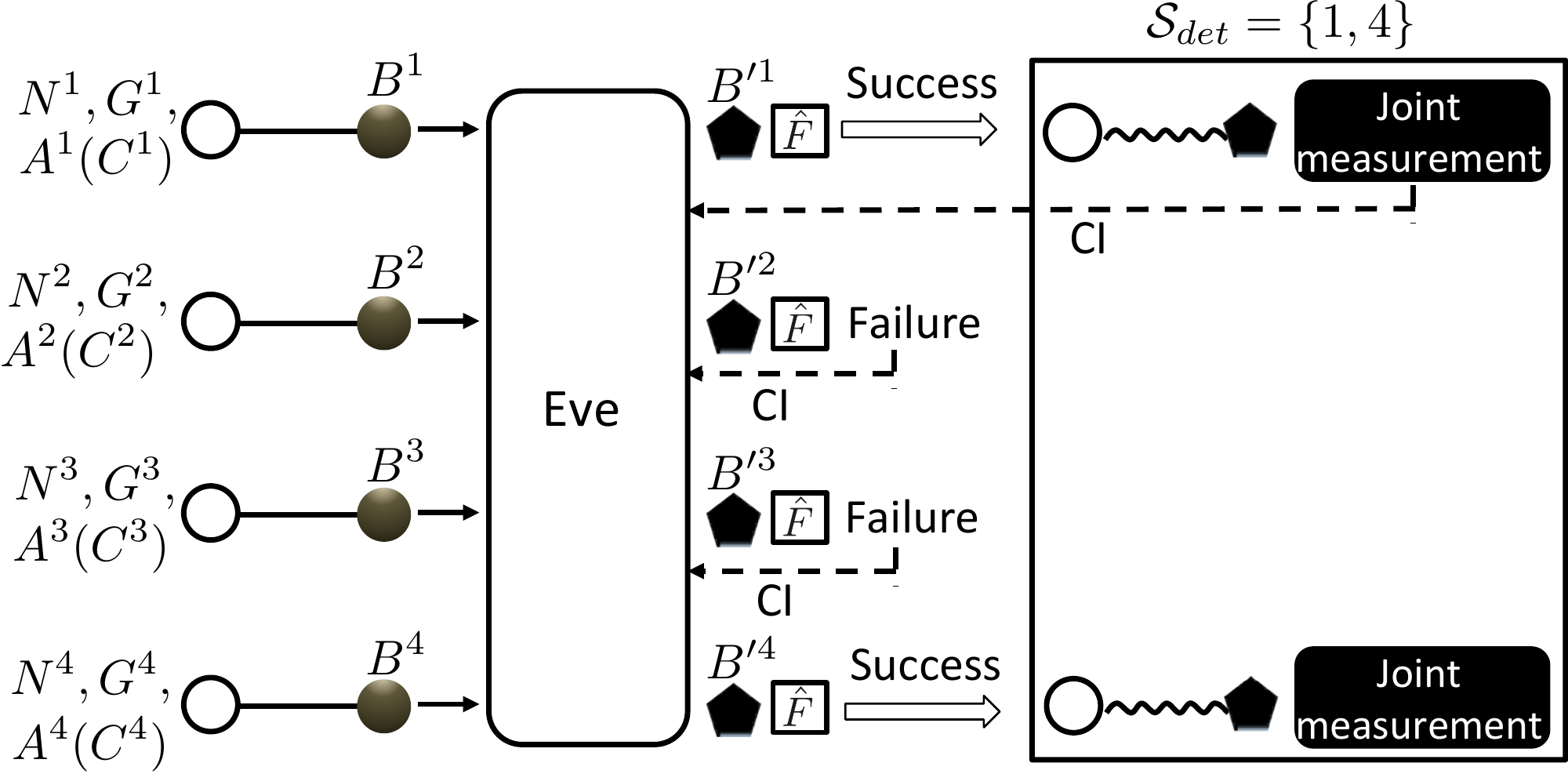}
\caption{Schematic representation of the virtual protocol for the case $N_{\U{sent}}=4$.
  For each $i$, Alice first sends the system $B^i$ to Bob via the quantum channel. 
  The eavesdropper performs an arbitrary operation on this system, and sends the system $B'^i$ to Bob.
  Bob applies to the systems received the filter operation $\hat{F}$, and we classify the events into two cases depending on whether this filter operation is
  successful (the set of such successful events is denoted by $\mathcal{S}_{\U{det}}$, in the example $\mathcal{S}_{\U{det}}=\{1,4\}$)
  or not. For the successful events, Alice and Bob perform a joint measurement on the systems $N^i,G^i,A^i(C^i)$ and $B'^i$
  depending if the outcomes of $N^i$ and $G^i$ are $n^i=1$ and $g^i\in\mathcal{G}^i_{\U{unt}}$
  ($n^i\neq 1$ or $g^i\notin\mathcal{G}^i_{\U{unt}}$), respectively.
  For the failure events, Alice performs a measurement on the systems $N^i,G^i$ and $C^i$. 
  In any case, once they have performed their measurements, Alice and Bob announce the relevant 
  classical information (CI) over an authenticated public channel. This is illustrated with a dashed line in the figure.
  We can use this virtual protocol to prove the security of the actual protocol because the quantum and
  classical information that is accessible to Eve coincide in both cases (see Proposition $\ref{ProVirtual}$).
}
\label{fig:virtualrep}
          \end{figure}
In this Appendix, we derive an upper bound on the phase error rate $e_{\U{ph}|Z,Z,1,u,\U{det}}=N_{\U{ph},Z,Z,1,u,\U{det}}/S_{Z,Z,1,u,\U{det}}$.
Importantly, the procedure introduced below is valid against coherent attacks. 
            To derive an upper bound on $e_{\U{ph}|Z,Z,1,u,\U{det}}$, 
            we need to obtain an upper bound on the number of phase errors $N_{\U{ph},Z,Z,1,u,\U{det}}$.
            Below, 
            in the first subsection,
            we explain the derivation of $N_{\U{ph},Z,Z,1,u,\U{det}}$ with the restricted phase intervals in Eq.~(\ref{simpleRph}), 
            and in the second subsection, we explain how to obtain $N_{\U{ph},Z,Z,1,u,\U{det}}$ with the general phase intervals 
            in Eq.~(\ref{Rph}) from that with the restricted phase intervals.
            
            \subsection{Derivation of $N_{\U{ph},Z,Z,1,u,\U{det}}$ with restricted phase intervals in Eq.~(\ref{simpleRph})}
            \label{apEsimpleRph}
            First, from Eq.~(\ref{sendingrho}), we have that the total $N_{\U{sent}}$ states can be expressed as 
\begin{align}
  \sum_{\bm{g}^{N_{\U{sent}}}}\sqrt{p(\bm{g}^{N_{\U{sent}}})}\ket{\bm{g}^{N_{\U{sent}}}}_{\bm{G}^{N_{\U{sent}}}}\bigotimes^{N_{\U{sent}}}_{i=1}\ket{\Psi^{i}_{g^i}}_{C^iK^iN^iB^i}
  \ket{\theta^{i}_{c^{i},g^i},\mu^{i}_{k^{i},g^i}}_{\Theta^iM^i}
  \label{Rhototal}
\end{align}
with
\begin{align}
  \ket{\Psi^{i}_{g^i}}_{C^iK^iN^iB^i}:=\sum_{c^{i},k^{i}}\sqrt{p(c^{i})p(k^{i})}\ket{c^{i},k^{i}}_{C^iK^i}
  \sum_{n^{i}}\sqrt{p(n^{i}|\mu^{i}_{k^{i},g^i})}\ket{n^{i}}_{N^i}\ket{\hat{\Upsilon}^{i}(\theta^{i}_{c^{i},g^i},n^{i})}_{B^i}.
  \label{rhototal}
\end{align}
In particular, $\ket{\hat{\Upsilon}^{i}(\theta^i_{c^i,g^i},1)}_{B^i}$ can be expressed as
\begin{align}
  \ket{\hat{\Upsilon}^{i}(\theta^i_{c^i,g^i},1)}_{B^i}=
  (\ket{1}_{R^i}\ket{0}_{S^i}+e^{\U{i}\theta^i_{c^i,g^i}}\ket{0}_{R^i}\ket{1}_{S^i})/\sqrt{2},
    \label{qubit}
\end{align}
and we define the eigenstates of the $Y$ basis as $\ket{Y}_{B^i}:=\ket{1}_{R^i}\ket{0}_{S^i}$
and $\ket{Y^{\perp}}_{B^i}:=\ket{0}_{R^i}\ket{1}_{S^i}$,
the ones of the $Z$ basis are $\ket{Z}:=(\ket{Y}+\ket{Y^{\perp}})/\sqrt{2}$ 
and $\ket{Z^{\perp}}:=(-\U{i}\ket{Y}+\U{i}\ket{Y^{\perp}})/\sqrt{2}$,
and the ones of the $X$ basis are $\ket{X}:=(\ket{Z}+\ket{Z^{\perp}})/\sqrt{2}$ and $\ket{X^{\perp}}:=(\ket{Z}-\ket{Z^{\perp}})/\sqrt{2}$.
This means that the state $\ket{\hat{\Upsilon}^{i}(\theta^i_{c^i,g^i},1)}_{B^i}$ can be rewritten in terms of the $Z$
eigenstates as
$\ket{\hat{\Upsilon}^{i}(\theta^i_{c^i,g^i},1)}_{B^i}=e^{\U{i}\theta^i_{c^i,g^i}/2}(\cos\frac{\theta^i_{c^i,g^i}}{2}
\ket{Z}_{B^i}+\sin\frac{\theta^i_{c^i,g^i}}{2}\ket{Z^{\perp}}_{B^i})$. 

For convenience, we describe Alice's state preparation process of $\ket{\Psi^{i}_{g^i}}_{C^iK^iN^iB^i}$ for $g^i\in\mathcal{G}^i_{\U{unt}}$
and $n^i=1$ by means of an entanglement-based scheme. That is, we assume that she first generates the state
\begin{align}
  \ket{\Psi^{i}_{g^i,1}}_{C^iK^iB^i}:=\sum_{k^{i}}\sqrt{p(k^{i})p(1|\mu^i_{k^i,g^i})}\ket{k^{i}}_{K^i}
  \left[\sqrt{p^{A}_Z}\ket{\psi^{i}_Z(\theta^i_{0_Z,g^i},\theta^i_{1_Z,g^i})}_{C^iB^i}+
\sqrt{p^{A}_X}e^{-\U{i}\theta^i_{0_X,g^i}/2}\ket{0_X}_{C^i}\ket{\hat{\Upsilon}^{i}(\theta^i_{0_X,g^i},1)}_{B^i}\right]
  \label{n1untagZ}
  \end{align}
with $\ket{\psi^{i}_{Z}(\theta^i_{0_Z,g^i},\theta^i_{1_Z,g^i})}_{C^iB^i}:
=\sum^1_{x=0}e^{-\U{i}\theta^i_{x_Z,g^i}/2}\ket{x_Z}_{C^i}\ket{\hat{\Upsilon}^{i}(\theta^i_{x_Z,g^i},1)}_{B^i}/\sqrt{2}$,
and afterwards she measures the system $C^i$ to obtain $c^i$.
In this subsection, 
given $c^{i}=c$ and $g^i\in\mathcal{G}^i_{\U{unt}}$, we suppose that $\theta^{i}_{c,g^i}$ lies in the following interval for any $i$: 
\begin{align}
\label{theta}
\theta^{i}_{0_Z,g^i}\in R^{0_Z}_{\U{\U{ph}}}=[-\theta,\theta],~~
\theta^{i}_{1_Z,g^i}\in R^{1_Z}_{\U{\U{ph}}}=[\pi-\theta,\pi+\theta],~~
\theta^{i}_{0_X,g^i}\in R^{0_X}_{\U{\U{ph}}}=\left[\frac{\pi}{2}-\theta,\frac{\pi}{2}+\theta\right]~~~\left(\U{with}~0\le\theta<\frac{\pi}{6}\right).
\end{align}
To prove the security of the key generated from the $Z$ basis, we follow the definition of the phase error rate.
That is, we consider Alice and Bob's hypothetical measurements on their systems $C^i$ and $B'^i$
(where $B'^i$ represents the system $B^i$ after Eve's intervention) in the $X$ basis 
given that Alice prepared the state $\ket{\psi^{i}_Z(\theta^i_{0_Z,g^i},\theta^i_{1_Z,g^i})}_{C^iB^i}$.
In this virtual scenario, we have therefore, that Alice sends Bob the state $\ket{\psi^i_{\U{vir},m,g^i}}_{B^i}$ with probability
$p_{\U{vir},m,g^i}^{i}$ by projecting the system $C^i$ in the basis $\{\ket{+}_{C^i},\ket{-}_{C^i}\}$ with
$\ket{\pm}_{C^i}:=(\ket{0_Z}_{C^i}\pm\ket{1_Z}_{C^i})/\sqrt{2}$, where
\begin{align} 
\label{VirState}
\hat{P}[\ket{\psi^i_{\U{vir},m,g^i}}_{B^i}]
=&\frac{\U{tr}_{C^i}
  \left[\hat{P}[\ket{m}_{C^i}]\hat{P}[\ket{\psi_Z^{i}(\theta^i_{0_Z,g^i},\theta^i_{1_Z,g^i})}_{C^iB^i}]\right]}{p_{\U{vir},m,g^i}^{i}}
\notag\\
=&\begin{cases}
\hat{P}[\ket{\hat{\Upsilon}^i(\frac{\theta^i_{0_Z,g^i}+\theta^i_{1_Z,g^i}}{2},1)}_{B^i}]
& (\text{$m=+$}),\\
\hat{P}[\ket{\hat{\Upsilon}^i(\frac{\theta^i_{0_Z,g^i}+\theta^i_{1_Z,g^i}}{2}+\pi,1)}_{B^i}]
& (\text{$m=-$}),
\end{cases}
\end{align}
and 
\begin{align} 
\label{VirPro}
p_{\U{vir},\pm,g^i}^{i}={\rm tr}\left[\hat{P}[\ket{\pm}_{C^i}]\hat{P}[\ket{\psi_Z^{i}(\theta^i_{0_Z,g^i},\theta^i_{1_Z,g^i})}_{C^iB^i}]\right]
=\frac{1\pm\cos\frac{\theta^{i}_{0_Z,g^i}-\theta^{i}_{1_Z,g^i}}{2}}{2}.
\end{align}
For later convenience, we define 
\begin{align}
  \theta^i_{0_{\U{vir}},g^i}:=\frac{\theta^i_{0_Z,g^i}+\theta^i_{1_Z,g^i}}{2},
  ~~\theta^i_{1_{\U{vir}},g^i}:=\frac{\theta^i_{0_Z,g^i}+\theta^i_{1_Z,g^i}}{2}+\pi.
  \label{VirStatepro}
\end{align}
Thanks to Proposition~$\ref{ProVirtual}$ below, we are allowed to evaluate the security of the actual protocol
by using the virtual scheme illustrated in Fig.~\ref{fig:virtualrep}.
In particular, we consider a virtual protocol where for each signal emission, Alice can in principle
prepare the following state if $g^i\in\mathcal{G}^i_{\U{unt}}$ and $n^i=1$ instead of $\ket{\Psi^i_{g^i,1}}_{C^iK^iB^i}$ as
\begin{align}
  \label{virtualstate}
  \ket{\Psi^i_{g^i,1,\U{vir}}}_{A^iK^iB^i}=
  \sum_{k^{i}}\sqrt{p(k^{i})p(1|\mu^i_{k^i,g^i})}\ket{k^{i}}_{K^i}\left[\sum^4_{\alpha^{i}=0}\sqrt{p^i_{\alpha^{i},g^i}}\ket{\alpha^{i}}_{A^i}\ket{\psi^{i}_{\alpha^{i},g^i}}_{B^i}\right],
\end{align}
where the system $A^i$ is stored by Alice in a quantum memory.
Here, the states $\{\ket{\psi_{\alpha^{i},g^i}^{i}}\}_{\alpha^{i}}$ are defined as
\begin{align}
\ket{\psi_{0,g^i}^{i}}=\ket{\psi^i_{\U{vir},+,g^i}},&~\ket{\psi_{1,g^i}^{i}}=\ket{\psi^i_{\U{vir},-,g^i}},\notag\\
\ket{\psi_{2,g^i}^{i}}=\ket{\hat{\Upsilon}(\theta^i_{0_Z,g^i},1)},~\ket{\psi_{3,g^i}^{i}}&=\ket{\hat{\Upsilon}(\theta_{1_Z,g^i},1)}~\U{and}~
\ket{\psi_{4,g^i}^{i}}=\ket{\hat{\Upsilon}(\theta_{0_X,g^i},1)},
\end{align}
and the probabilities
$\{p_{\alpha^{i},g^i}^{i}\}_{\alpha^{i}}$ are given by
\begin{align}
  p_{0,g^i}^{i}=p^{A}_Zp^{B}_Zp_{\U{vir},+,g^i}^{i},~p_{1,g^i}^{i}=p^{A}_Zp^{B}_Zp_{\U{vir},-,g^i}^{i},~
  p_{2,g^i}^{i}=p_{3,g^i}^{i}=p^{A}_Zp^{B}_X/2~\U{and}~p_{4,g^i}^{i}=p^{A}_X.
  \label{probabilityLT}
  \end{align}
Alice sends Bob the system $B^i$ in Eq.~(\ref{Rhototal}), and Eve performs an arbitrary operation on this system.
On the receiving side, thanks to the assumption (B-1) in Sec.~\ref{assbob}, Bob can apply the filter operation
$\hat{F}$ to the system $B'^i$ to determine 
whether he will have a detection event ($y^i\neq \emptyset$) or not ($y^i=\emptyset$) prior to selecting the measurement basis, which is represented by the
Kraus operators $\left\{\sqrt{\hat{I}_{B'^i}-\hat{M}_{\emptyset}},\sqrt{\hat{M}_{\emptyset}}\right\}$.
Depending on this result, we classify all the emissions into two types of events:
\begin{align}
\U{  (i)~a~successful~event~(which~is~associated~to~the~operator}~\sqrt{\hat{I}_{B'^i}-\hat{M}_{\emptyset}}),\notag
\end{align}
  and
\begin{align}
  \U{(ii)~a~failure~event~(which~is~associated~to~the~operator}~\sqrt{\hat{M}_{\emptyset}}).\notag
\end{align}
We denote the set of successful events by $\mathcal{S}_{\U{det}}:=\{i|y^{i}\neq\emptyset\}$ where $S_{\U{det}}:=|\mathcal{S}_{\U{det}}|$. 
Afterwards, we consider that Alice and Bob perform joint measurements on all the signals. 
In particular, for the successful event identified above as 
(i), we suppose that Alice measures her systems $N^i$ and $G^i$, and if the outcomes are $n^i=1$ and $g^i\in\mathcal{G}^i_{\U{unt}}$, 
she measures the systems $A^i$ to obtain the outcome $\alpha^i\in\{0,1,2,3,4\}$.
If $\alpha^i\in\{0,1,2,3\}$, then Bob measures the system $B'^i$ in the $X$ basis to obtain the outcome $y^i\in\{0,1\}$, and
if $\alpha^i=4$, then Bob measures the system $B'^i$ in the $Z~(X)$ basis
with probability $p^B_Z~(p^B_X)$ and obtains the outcome $y^i\in\{0,1\}$. 
Also, among the successful events of the type (i),
if $n^i\neq1$ or $g^i\notin\mathcal{G}^i_{\U{unt}}$, Alice and Bob measure their systems $C^i$ and $B'^i$ and obtain the 
outcomes $c^i\in\mathcal{C}$ and $y^i\in\{0,1\}$.

For the failure event identified above as (ii), Alice measures her systems $N^i$, $G^i$ and $C^i$ and obtains the outcomes
$n^i\in[0,\infty)$, $g^i\in\mathcal{G}^i_{\U{unt}}\cup\overline{\mathcal{G}^i_{\U{unt}}}$ and $c^i\in\mathcal{C}$, respectively. 
    Importantly, it can be shown that, from Eve's perspective, this virtual protocol is completely equivalent to the actual protocol
    given that Alice and Bob's announcements and classical post-processing are chosen appropriately.

  \begin{pro}
  \label{ProVirtual}
  From Eve's perspective, the virtual protocol described above is equivalent to the actual protocol, namely,
  the quantum and classical information available to Eve are exactly the same for both protocols.
  Also, the correspondence between classical information $(c^i,k^i)$ and the quantum state
  in the actual protocol is identical to the one of the virtual protocol.
\end{pro}
{\bf Proof of Proposition~\ref{ProVirtual}}. First, we show that Eve's accessible quantum information is the same in both protocols.
The case of the tagged signal ($t^i=t$) or non-single-photon ($n^i\neq1$) emissions is clear, as the quantum states sent by Alice in both schemes coincide. 
Hence, the quantum information that is available to Eve is obviously the same in this case. 
As for the untagged single-photon emissions, we have that in the virtual protocol Alice sends Bob the virtual states
[see Eq.~(\ref{virtualstate})], while in the actual protocol, she does not.
Still, we have that also in this case Eve's accessible quantum information of the system $B^i$ is the same between the two protocols 
because the following relation
\begin{align}
  {\rm tr}_{A^iK^i}\hat{P}[\ket{\Psi^i_{g^i,1,\U{vir}}}_{A^iK^iB^i}]={\rm tr}_{C^iK^i}\hat{P}[\ket{\Psi^i_{g^i,1}}_{C^iK^iB^i}]
  \label{equiv}
\end{align}
holds. To conclude, we show that Eve's accessible classical information is also equal in both protocols. 
  For the case of the tagged signal ($t^i=t$) or non-single-photon ($n^i\neq1$) emission event,
  since the quantum states are the same in both the actual and virtual protocols,
  the classical information declared by Alice and Bob in both schemes coincide. 
  Hence, we only need to consider the equivalence between both protocols for the case of $n^i=1$ and $t^i=u$. 
  For this, we suppose that after finishing each of their joint measurements, Alice and Bob announce the classical information
  as follows.
Whenever their measurement outcome on the system $A^i$ is $\alpha^{i}\in\{0,1\}$, Alice and Bob declare over an authenticated public channel the $Z$ basis.
Also, Alice (Bob) announces the $Z$ basis ($X$ basis and its measurement outcome) when
$\alpha^{i}\in\{2,3\}$.
Finally, if $\alpha^{i}=4$ and Bob's basis choice is $Z$ ($X$),
Alice announces the $X$ basis and Bob announces the $Z$ ($X$) basis and its measurement outcome. 
This way, it is easy to see that the classical information announced in the actual and virtual protocols coincide.
Furthermore, due to Eq.~(\ref{rhototal}), we can confirm that the correspondence between the classical information $(c^i,k^i)$ and the quantum state in the actual protocol is equivalent to the one of the virtual protocol. 
This ends the proof of the Proposition.\sq 

Thanks to this Proposition, the security statements that we derive for the virtual protocol can be applied as well to
the actual protocol.

Next, we describe the state of the $i^{\U{th}}$ systems $G^i,C^i,K^i,N^i$ and $B'^i$ from Alice and Bob's point of view. 
Given that Bob has finished measuring the $(i-1)^{\U{th}}$ incoming pulse, Alice and Bob obtained the measurement outcomes
${\bm \zeta}^{i-1}:=(\bm{n}^{i-1},\bm{g}^{i-1},\bm{\gamma}^{i-1},\bm{y}^{i-1})$ with $\gamma^i$ being $\alpha^i$ or $c^i$
depending on whether the measured system is $A^i$ or $C^i$.
Once the outcome ${\bm \zeta}^{i-1}$ and $\bm{g}^{\ge i}$ are fixed, the operation acting on the system $B^i$ is mathematically modeled by 
the completely positive (CP) map $\Lambda^{i}_{{\bm \zeta}^{i-1},\bm{I}^{i-1},\bm{g}^{\ge i}}$ that 
includes Eve's operation depending on the iterative information $\bm{I}^{i-1}$ announced by Alice and Bob up to the $(i-1)^{\U{th}}$
trials. With this CP map, the state in the $i^{\U{th}}$ systems $G^i,C^i,K^i,N^i$ and $B'^i$ is of the form
\begin{align}
\frac{\hat{\sigma}_{{\bm \zeta}^{i-1},\bm{I}^{i-1}}}{\U{tr}[\hat{\sigma}_{{\bm \zeta}^{i-1},\bm{I}^{i-1}}]},
  \end{align}
where $\hat{\sigma}_{{\bm \zeta}^{i-1},\bm{I}^{i-1}}$ is defined as
  \begin{align}
    \label{cptp}
    \hat{\sigma}_{{\bm \zeta}^{i-1},\bm{I}^{i-1}}:
    =&\sum_{\bm{g}^{\ge i+1}}p(\bm{g}^{i-1},\bm{g}^{\ge i+1})\hat{P}\left[\sum_{g^i}\sqrt{p(g^i|\bm{g}^{i-1},\bm{g}^{\ge i+1})}
      \Lambda^{i}_{{\bm \zeta}^{i-1},\bm{I}^{i-1},\bm{g}^{\ge i}}(\ket{g^i}_{G^i}\ket{\Psi^i_{g^i}}_{C^i,K^i,N^i,B^i})\right]\notag\\
    =&\sum_{s^i}\sum_{\bm{g}^{\ge i+1}}p(\bm{g}^{i-1},\bm{g}^{\ge i+1})\hat{P}\left[\sum_{g^i}\sqrt{p(g^i|\bm{g}^{i-1},\bm{g}^{\ge i+1})}
      \hat{A}^{i}_{s^i,{\bm \zeta}^{i-1},\bm{I}^{i-1},\bm{g}^{\ge i}}\ket{g^i}_{G^i}\ket{\Psi^i_{g^i}}_{C^i,K^i,N^i,B^i}\right].
  \end{align}
  Here, the set of Kraus operators $\{\hat{A}^{i}_{s^i,{\bm \zeta}^{i-1},\bm{I}^{i-1},\bm{g}^{\ge i}}\}_{s^i}$ acting on the system $B^i$ 
  depends on the measurement results $\bm{\zeta}^{i-1}$ and the iterative information $\bm{I}^{i-1}$ announced by Alice and Bob.
  
Let us now derive an upper bound on the number of phase errors $N_{\U{ph},Z,Z,1,u,\U{det}}$. 
For this, we consider the joint measurement performed by Alice and Bob on the $i^{\U{th}}$ event with $i\in \mathcal{S}_{\U{det}}$.
In particular, we are interested in the probability of obtaining the measurement outcome $n^{i}=1$ on the system $N^i$,
the measurement outcome $g^{i}\in\mathcal{G}^i_{\U{unt}}$ (namely, $t^i=u$) on the system $G^i$,
the measurement outcome $\alpha^{i}=\alpha\in\{0,1,2,3,4\}$ on the system $A^i$, 
and the measurement outcome $y^{i}=y\in\{0,1\}$ in the $X$ basis measurement on the system $B'^i$ given that
Bob's filter operation $\hat{F}$ was successful.
Such probability can be expressed as
\begin{align}
  &p(n^{i}=1,t^{i}=u,\alpha^{i}=\alpha,b^i=X,y^{i}=y|\bm{\zeta}^{i-1},y^{i}\neq\emptyset)\notag\\
  =&\U{tr}\Bigg\{
  \sum_{g^i\in\mathcal{G}^i_{\U{unt}}}
  \hat{P}[\ket{1}_{N^i}\ket{g^i}_{G^i}\ket{\alpha}_{A^i}]\otimes q_{\alpha}\hat{M}_{y,X}
  \frac{\hat{\sigma}_{{\bm \zeta}^{i-1},\bm{I}^{i-1}}}
       {\U{tr}[\mathcal{E}^{i}(\hat{\sigma}_{{\bm \zeta}^{i-1},\bm{I}^{i-1}})]}\Bigg\}\notag\\
       =&\sum_{g^i\in\mathcal{G}^i_{\U{unt}}}\sum_{\bm{g}^{\ge i+1}}p(\bm{g}^{N_{\U{sent}}})\sum_{k^{i}}p(k^{i})p(1|\mu_{k^{i},g^i})
       \frac{
        \U{tr}\Big[q_{\alpha}p^{i}_{\alpha,g^i}\sum_{s^i}
    (\hat{A}^{i}_{s^i,{\bm \zeta}^{i-1},\bm{I}^{i-1},\bm{g}^{\ge i}})^{\dagger}\hat{M}_{y,X}
    (\hat{A}^{i}_{s^i,{\bm \zeta}^{i-1},\bm{I}^{i-1},\bm{g}^{\ge i}})
    \hat{P}[\ket{\psi^{i}_{\alpha,g^i}}_{B^i}\Big]}
          {\U{tr}[\mathcal{E}^{i}(\hat{\sigma}_{{\bm \zeta}^{i-1},\bm{I}^{i-1}})]}\notag\\      
      =&\left<C^i_{g^i}q_{\alpha}p^{i}_{\alpha,g^i}
      \U{tr}\Big[\hat{T}^{i}_{y,X,{\bm \zeta}^{i-1},\bm{I}^{i-1},\bm{g}^{\ge i}}\hat{P}[\ket{\psi^{i}_{\alpha,g^i}}_{B^i}]\Big]
      \right>.
  \label{Pr}
\end{align}
Here, we define 
$\mathcal{E}^{i}(\hat{\rho}):=\sqrt{\hat{I}_{B'^i}-\hat{M}_{\emptyset}}\hat{\rho}\sqrt{\hat{I}_{B'^i}-\hat{M}_{\emptyset}}^{\dagger}$,
$C^i_{g^i}:=\sum_{k^{i}}p(k^{i})p(1|\mu_{k^{i},g^i})$, $q_{\alpha}:=1$ for $\alpha\in\{0,1,2,3\}$ and 
$q_4:=p^B_X$. Also, we define the operator $\hat{T}^{i}_{y,X,{\bm \zeta}^{i-1},\bm{I}^{i-1},\bm{g}^{\ge i}}$ as
    \begin{align}
\hat{T}^{i}_{y,X,{\bm \zeta}^{i-1},\bm{I}^{i-1},\bm{g}^{\ge i}}:=\frac{
    \sum_{s^i}
    (\hat{A}^{i}_{s^i,{\bm \zeta}^{i-1},\bm{I}^{i-1},\bm{g}^{\ge i}})^{\dagger}\hat{M}_{y,X}
    (\hat{A}^{i}_{s^i,{\bm \zeta}^{i-1},\bm{I}^{i-1},\bm{g}^{\ge i}})}
    {\U{tr}[\mathcal{E}^{i}(\hat{\sigma}_{{\bm \zeta}^{i-1},\bm{I}^{i-1}})]}.
          \end{align}
    In Eq.~(\ref{Pr}), for convenience of notation, we have defined 
    $\left<A\right>:=\sum_{g^i\in\mathcal{G}^i_{\U{unt}}}\sum_{\bm{g}^{\ge i+1}}p(\bm{g}^{N_{\U{sent}}})A$. 

    If we define $N_{1,u,\alpha,y,X}$ as the number of events where Alice obtains the measurement outcome $n^{i}=1$ on the system $N^i$,
    $g^i\in\mathcal{G}^i_{\U{unt}}$ on the system $G^i$, $\alpha^{i}=\alpha\in\{0,1\}$ on her system $A^i$ and Bob obtains the outcome 
    $y^{i}=y\in\{0,1\}$ in the $X$ basis measurement on the system $B'^i$, 
    the phase error rate is defined as the rate at which Bob obtains the outcome $y^{i}=\alpha\oplus1$ on the system $B'^i$
    among $\sum_{\alpha=0,1,y=0,1}N_{1,u,\alpha,y,X}$.
    The phase error rate for the untagged single-photon emissions can be expressed as
\begin{align}
  e_{\U{ph}|Z,Z,1,u,\U{det}}=\frac{\sum^1_{\alpha=0}N_{1,u,\alpha,y=\alpha\oplus1,X}}{\sum^1_{\alpha=0}\sum^1_{y=0}N_{1,u,\alpha,y,X}}
  =:\frac{N_{\U{ph},Z,Z,1,u,\U{det}}}{\sum^1_{\alpha=0}\sum^1_{y=0}N_{1,u,\alpha,y,X}}
  =\frac{N_{\U{ph},Z,Z,1,u,\U{det}}}{S_{Z,Z,1,u,\U{det}}}.
  \label{phaseerror}
\end{align}
In the third equality, we use the assumption (B-1) in Sec.~\ref{assbob} which states that the efficiency of Bob's measurement
is the same for the bases $Z$ and $X$. In Eq.~(\ref{phaseerror}), the denominator can be estimated directly with the decoy-state method.
Next, we calculate an upper bound for the quantity $N_{\U{ph},Z,Z,1,u,\U{det}}$. For this, we first relate 
$N_{\U{ph},Z,Z,1,u,\U{det}}$ with the sum of the probabilities
$\sum^1_{\alpha=0}p(n^{i}=1,t^{i}=u,\alpha^{i}=\alpha,b^i=X,y^{i}=\alpha\oplus1|\bm{\zeta}^{i-1},y^{i}\neq\emptyset)$ over 
$i\in \mathcal{S}_{\U{det}}$. This can be done by using Azuma's inequality~\cite{Azuma1967}; we obtain
\begin{align}
  p\left( N_{\U{ph},Z,Z,1,u,\U{det}}-\sum_{i\in \mathcal{S}_{\U{det}}}\sum^1_{\alpha=0}
    p(n^{i}=1,t^{i}=u,\alpha^{i}=\alpha,b^i=X,y^{i}=\alpha\oplus1|\bm{\zeta}^{i-1},y^{i}\neq\emptyset)
    \ge g_{\U{A}}(N_{\U{det}},\epsilon^{\U{ph},Z,1,u}_{\U{A}})\right)\le\epsilon^{\U{ph},Z,1,u}_{\U{A}},
  \label{azuma1}
\end{align}
where $g_{\U{A}}(x,y):=\sqrt{2x\ln1/y}$.
From Eqs.~(\ref{Pr}) and (\ref{azuma1}), we have that $N_{\U{ph},Z,Z,1,u,\U{det}}$ can be upper-bounded by
\begin{align}
N_{\U{ph},Z,Z,1,u,\U{det}}\le\sum_{i\in \mathcal{S}_{\U{det}}}\sum^1_{\alpha=0}
  \left<C^i_{g^i}p^{i}_{\alpha,g^i}
      \U{tr}\Big[\hat{T}^{i}_{\alpha\oplus1,X,{\bm \zeta}^{i-1},\bm{I}^{i-1},\bm{g}^{\ge i}}\hat{P}[\ket{\psi^{i}_{\alpha,g^i}}_{B^i}]\Big]
      \right>+g_{\U{A}}(N_{\U{det}},\epsilon^{\U{ph},Z,1,u}_{\U{A}})
  \label{a2}
\end{align}
except for error probability $\epsilon^{\U{ph},Z,1,u}_{\U{A}}$. The first term on the rhs represents the transmission rate of the
virtual states $\ket{\psi^i_{0,g^i}}_{B^i}$ and $\ket{\psi^i_{1,g^i}}_{B^i}$. To upper bound this quantity using the experimentally
available data, we express ${\rm tr}\Big[\hat{T}^{i}_{\alpha\oplus1,X,{\bm \zeta}^{i-1},\bm{I}^{i-1},\bm{g}^{\ge i}}
  \hat{P}[\ket{\hat{\Upsilon}(\theta^i_{c,g^i},1)}_{B^i}]\Big]$ 
for $g^i\in\mathcal{G}^i_{\U{unt}}$ and $c\in\mathcal{C}$, in terms of the transmission rates of the Pauli operators $\hat{I}_{B^i}, \hat{X}_{B^i}$ and
$\hat{Z}_{B^i}$. In particular, since $\hat{P}[\ket{\hat{\Upsilon}(\theta,1)}_{B^i}]$ can be decomposed as
$\hat{P}[\ket{\hat{\Upsilon}(\theta,1)}_{B^i}]=(\hat{I}_{B^i}+\sin{\theta}\hat{X}_{B^i}+\cos{\theta}\hat{Z}_{B^i})/2$, we have that
\begin{equation}
  \label{relation}
  \begin{pmatrix}
      $tr$[\hat{T}^{i}_{\alpha\oplus1,X,{\bm \zeta}^{i-1},\bm{I}^{i-1},\bm{g}^{\ge i}}\hat{P}[\ket{\hat{\Upsilon}^i(\theta^i_{0_Z,g^i},1)}_{B^i}] \\
      $tr$[\hat{T}^{i}_{\alpha\oplus1,X,{\bm \zeta}^{i-1},\bm{I}^{i-1},\bm{g}^{\ge i}}\hat{P}[\ket{\hat{\Upsilon}^i(\theta^i_{1_Z,g^i},1)}_{B^i}] \\
      $tr$[\hat{T}^{i}_{\alpha\oplus1,X,{\bm \zeta}^{i-1},\bm{I}^{i-1},\bm{g}^{\ge i}}\hat{P}[\ket{\hat{\Upsilon}^i(\theta^i_{0_X,g^i},1)}_{B^i}]  
    \end{pmatrix}=M^{i}_{g^i}
 \begin{pmatrix}
      $tr$[\hat{T}^{i}_{\alpha\oplus1,X,{\bm \zeta}^{i-1},\bm{I}^{i-1},\bm{g}^{\ge i}}\hat{I}_{B^i}/2] \\
      $tr$[\hat{T}^{i}_{\alpha\oplus1,X,{\bm \zeta}^{i-1},\bm{I}^{i-1},\bm{g}^{\ge i}}\hat{X}_{B^i}/2] \\
      $tr$[\hat{T}^{i}_{\alpha\oplus1,X,{\bm \zeta}^{i-1},\bm{I}^{i-1},\bm{g}^{\ge i}}\hat{Z}_{B^i}/2] 
    \end{pmatrix},
\end{equation}
where $M^{i}_{g^i}:=(\vec{V}^{i}_{0_Z,g^i}, \vec{V}^{i}_{1_Z,g^i}, \vec{V}^{i}_{0_X,g^i})^{\U{T}}$ with
$(\vec{V}^{i}_{c,g^i})^{\U{T}}:=(1, \sin{\theta_{c,g^i}^{i}}, \cos{\theta_{c,g^i}^{i}})$, and where T represents the transpose operator. 
To calculate ${\rm tr}[\hat{T}^{i}_{\alpha\oplus1,X,{\bm \zeta}^{i-1},\bm{I}^{i-1},\bm{g}^{\ge i}}\hat{I}_{B^i}/2]$,
${\rm tr}[\hat{T}^{i}_{\alpha\oplus1,X,{\bm \zeta}^{i-1},\bm{I}^{i-1},\bm{g}^{\ge i}}\hat{X}_{B^i}/2]$, 
and ${\rm tr}[\hat{T}^{i}_{\alpha\oplus1,X,{\bm \zeta}^{i-1},\bm{I}^{i-1},\bm{g}^{\ge i}}\hat{Z}_{B^i}/2]$,
we use information from the states that are sent in the actual protocol.
In doing so, we can upper-bound the first term of the rhs in Eq.~(\ref{a2}) as
\begin{align}
  \label{transdecom}
  &
  \sum_{i\in \mathcal{S}_{\U{det}}}\sum^1_{\alpha=0}
  \left<C^i_{g^i}p^{i}_{\alpha,g^i}\U{tr}\Big[\hat{T}^{i}_{\alpha\oplus1,X,{\bm \zeta}^{i-1},\bm{I}^{i-1},\bm{g}^{\ge i}}
    \hat{P}[\ket{\psi^{i}_{\alpha,g^i}}_{B^i}]\Big]
  \right>
\notag\\
  \le& \sum^1_{\alpha=0}p_{\alpha}^{\U{U}}\sum_{i\in \mathcal{S}_{\U{det}}}\left<
  C^i_{g^i}\U{tr}\Big[\hat{T}^{i}_{\alpha\oplus1,X,{\bm \zeta}^{i-1},\bm{I}^{i-1},\bm{g}^{\ge i}}\hat{P}[\ket{\psi^{i}_{\alpha,g^i}}_{B^i}]\Big]
  \right>\notag \\
  =&\sum^1_{\alpha=0}p_{\alpha}^{\U{U}}\sum_{i\in \mathcal{S}_{\U{det}}}
  \left<C^i_{g^i}\U{tr}\Big[\hat{T}^{i}_{\alpha\oplus1,X,{\bm \zeta}^{i-1},\bm{I}^{i-1},\bm{g}^{\ge i}}
    \Big(\hat{I}_{B^i}/2+\sin{\theta_{{\alpha}_{\rm vir},g^i}^{i}}\hat{X}_{B^i}/2+\cos{\theta_{{\alpha}_{\rm vir},g^i}^{i}}\hat{Z}_{B^i}/2\Big)\Big]\right>
  \notag\\
=&\sum^1_{\alpha=0}p_{\alpha}^{\U{U}}\sum_{i\in \mathcal{S}_{\U{det}}}\left<C^i_{g^i}\sum_{c\in\mathcal{C}}\U{tr}\Big[\Gamma_{{\alpha},c,g^i}^{i}
    \hat{T}^{i}_{\alpha\oplus1,X,{\bm \zeta}^{i-1},\bm{I}^{i-1},\bm{g}^{\ge i}}
    \hat{P}[\ket{\hat{\Upsilon}(\theta^i_{c,g^i},1)}_{B^i}]\Big]\right>\nonumber\\ 
\le& \sum^1_{\alpha=0}p_{\alpha}^{\U{U}}\sum_{c\in\mathcal{C}}\Gamma_{{\alpha},c}^{\U{U}}\sum_{i\in \mathcal{S}_{\U{det}}}\left<C^i_{g^i}{\rm tr}
  \Big[\hat{T}^{i}_{\alpha\oplus1,X,{\bm \zeta}^{i-1},\bm{I}^{i-1},\bm{g}^{\ge i}}\hat{P}[\ket{\hat{\Upsilon}(\theta^i_{c,g^i},1)}_{B^i}]\Big]
  \right>\notag\\
=&\sum^1_{\alpha=0}p_{\alpha}^{\U{U}}\Big(\Gamma_{{\alpha}, 0_Z}^{\U{U}} \sum_{i\in \mathcal{S}_{\U{det}}}
p(n^{i}=1,t^{i}=u,\alpha^{i}=2,b^i=X,y^{i}=\alpha\oplus1|\bm{\zeta}^{i-1},y^{i}\neq\emptyset)/p^{i}_{2,g^i}\notag\\
+&\Gamma_{{\alpha}, 1_Z}^{\U{U}}\sum_{i\in \mathcal{S}_{\U{det}}}p(n^{i}=1,t^{i}=u,\alpha^{i}=3,b^i=X,y^{i}=\alpha\oplus1|\bm{\zeta}^{i-1},y^{i}
  \neq\emptyset)/p^{i}_{3,g^i}\notag\\
  +&\Gamma_{{\alpha}, 0_X}^{\U{U}}\sum_{i\in \mathcal{S}_{\U{det}}}
  p(n^{i}=1,t^{i}=u,\alpha^{i}=4,b^i=X,y^{i}=\alpha\oplus1|\bm{\zeta}^{i-1},y^{i}\neq\emptyset)/p^B_Xp^{i}_{4,g^i}\Big).
\end{align}
In the first inequality, the parameter $p^{\U{U}}_{\alpha}$ (with $\alpha\in\{0,1\})$ is an upper bound on
$p^{i}_{0(1),g^i}:=p^A_Zp^B_Zp_{{\rm vir},+(-),g^i}^{i}$, where $p_{{\rm vir},\pm,g^i}^{i}$ is defined in Eq.~(\ref{VirPro}), 
and these quantities are given by
\begin{align}
  p^{\U{U}}_0=p^{\U{U}}_1=p^A_Zp^B_Z(1+\sin\theta)/2.
  \label{palphaU}
  \end{align}
In the first equality in Eq.~(\ref{transdecom}), the parameters $\theta^i_{0_{\U{vir}},g^i}$ and $\theta^i_{1_{\U{vir}},g^i}$ are 
defined in Eq.~(\ref{VirStatepro}). 
The second equality in Eq.~(\ref{transdecom}) is due to Eq.~(\ref{relation}) and the parameters $\{\Gamma_{{\alpha},c,g^i}^{i}\}_{\alpha,c}$,
which connect the transmission rates of the Pauli operators to those of the actual states,
are respectively given by~\cite{PhysRevA.93.042325}
\begin{align}
\label{gs}
&\Gamma_{0,0_Z,g^i}^{i}(\theta_{0_Z,g^i}^{i},\theta_{1_Z,g^i}^{i},\theta_{0_X,g^i}^{i})\notag\\
=&\sin{\left(\frac{\theta_{0_Z,g^i}^{i}+\theta_{1_Z,g^i}^{i}
-2\theta_{0_X,g^i}^{i}}{4}\right)}\left[\sin{\left(\frac{\theta_{0_Z,g^i}^{i}+\theta_{1_Z,g^i}^{i}
      -2\theta_{0_X,g^i}^{i}}{4}\right)}-
  \sin{\left(\frac{-3\theta_{0_Z,g^i}^{i}+\theta_{1_Z,g^i}^{i}+2\theta_{0_X,g^i}^{i}}{4}\right)}\right]^{-1}, \notag \\
&\Gamma_{0,1_Z,g^i}^{i}(\theta_{0_Z,g^i}^{i},\theta_{1_Z,g^i}^{i},\theta_{0_X,g^i}^{i})\notag\\
=&\sin{\left(\frac{\theta_{0_Z,g^i}^{i}+\theta_{1_Z,g^i}^{i}
-2\theta_{0_X,g^i}^{i}}{4}\right)}\left[\sin{\left(\frac{\theta_{0_Z,g^i}^{i}+\theta_{1_Z,g^i}^{i}
-2\theta_{0_X,g^i}^{i}}{4}\right)}+\sin{\left(\frac{-\theta_{0_Z,g^i}^{i}+3\theta_{1_Z,g^i}^{i}
-2\theta_{0_X,g^i}^{i}}{4}\right)}\right]^{-1},\notag \\
&\Gamma_{0,0_X,g^i}^{i}(\theta_{0_Z,g^i}^{i},\theta_{1_Z,g^i}^{i},\theta_{0_X,g^i}^{i})\notag\\
=&\left[1-\cos{\left(\frac{\theta_{0_Z,g^i}^{i}-\theta_{1_Z,g^i}^{i}
}{2}\right)}\right]\left[\cos{\left(\frac{\theta_{0_Z,g^i}^{i}+\theta_{1_Z,g^i}^{i}
-2\theta_{0_X,g^i}^{i}}{2}\right)}-\cos{\left(\frac{\theta_{0_Z,g^i}^{i}-\theta_{1_Z,g^i}^{i}}{2}\right)}\right]^{-1}, \notag \\
&\Gamma_{1,0_Z,g^i}^{i}(\theta_{0_Z,g^i}^{i},\theta_{1_Z,g^i}^{i},\theta_{0_X,g^i}^{i})\notag\\
=&\cos{\left(\frac{\theta_{0_Z,g^i}^{i}+\theta_{1_Z,g^i}^{i}
-2\theta_{0_X,g^i}^{i}}{4}\right)}\left[\cos{\left(\frac{\theta_{0_Z,g^i}^{i}+\theta_{1_Z,g^i}^{i}
-2\theta_{0_X,g^i}^{i}}{4}\right)}-\cos{\left(\frac{-3\theta_{0_Z,g^i}^{i}+\theta_{1_Z,g^i}^{i}
+2\theta_{0_X,g^i}^{i}}{4}\right)}\right]^{-1}, \notag \\
&\Gamma_{1,1_Z,g^i}^{i}(\theta_{0_Z,g^i}^{i},\theta_{1_Z,g^i}^{i},\theta_{0_X,g^i}^{i})\notag\\
=&\cos{\left(\frac{\theta_{0_Z,g^i}^{i}+\theta_{1_Z,g^i}^{i}
-2\theta_{0_X,g^i}^{i}}{4}\right)}\left[\cos{\left(\frac{\theta_{0_Z,g^i}^{i}+\theta_{1_Z,g^i}^{i}
-2\theta_{0_X,g^i}^{i}}{4}\right)}-\cos{\left(\frac{-\theta_{0_Z,g^i}^{i}+3\theta_{1_Z,g^i}^{i}
-2\theta_{0_X,g^i}^{i}}{4}\right)}\right]^{-1}, \notag \\
&\Gamma_{1,0_X,g^i}^{i}(\theta_{0_Z,g^i}^{i},\theta_{1_Z,g^i}^{i},\theta_{0_X,g^i}^{i})\notag\\
=&\left[-1-\cos{\left(\frac{\theta_{0_Z,g^i}^{i}-\theta_{1_Z,g^i}^{i}
}{2}\right)}\right]\left[\cos{\left(\frac{\theta_{0_Z,g^i}^{i}+\theta_{1_Z,g^i}^{i}
-2\theta_{0_X,g^i}^{i}}{2}\right)}-\cos{\left(\frac{\theta_{0_Z,g^i}^{i}-\theta_{1_Z,g^i}^{i}}{2}\right)}\right]^{-1}.
\end{align}
In addition, the parameter $\Gamma^{\U{U}}_{\alpha, c}$ that appears in the second inequality within Eq.~(\ref{transdecom}) represents an upper
bound on $\Gamma_{\alpha,c,g^i}^{i}$ in the range of $g^i\in\mathcal{G}^i_{\U{unt}}$.
If the conditions stated in Eq.~(\ref{theta}) are satisfied, we have that these quantities have the form~\cite{PhysRevA.93.042325}: 
$\Gamma_{0,0_Z}^{\U{U}}=\frac{\sin{\theta}}{\sin{\theta}+\cos{\frac{3}{2}\theta}}$, $\Gamma_{0,1_Z}^{\U{U}}=\Gamma_{0,0_Z}^{\U{U}}$, 
$\Gamma_{0,0_X}^{\U{U}}=\frac{1-\sin{\theta}}{\cos{2\theta}-\sin{\theta}}$, 
$\Gamma_{1,0_Z}^{\U{U}}=\frac{\cos{\theta}}{\cos{\theta}-\sin{\frac{3}{2}\theta}}$, $\Gamma_{1,1_Z}^{\U{U}}=\Gamma_{1,0_Z}^{\U{U}}$,
and $\Gamma_{1,0_X}^{\U{U}}=-\frac{1-\sin{\theta}}{1+\sin{\theta}}$.
Finally, the last equality in Eq.~(\ref{transdecom}) is due to Eq.~(\ref{Pr}), where
we note that $p^i_{2,g^i},p^i_{3,g^i}$ and $p^i_{4,g^i}$ are constant values [see Eq.~(\ref{probabilityLT})]. 

The next step is to relate the sum of the probabilities 
$\sum_{i\in \mathcal{S}_{\U{det}}}p(n^{i}=1,t^{i}=u,\alpha^{i}=\beta,b^i=X,y^{i}=\alpha\oplus1|{\bm \zeta}^{i-1},y^{i}\neq\emptyset)$
with $\beta\in\{2,3,4\}$ that appear in Eq.~(\ref{transdecom}) to the actual number of such events, which we denote by 
$S_{c,1,u,\U{det},\alpha\oplus1,X}$ with $c\in\mathcal{C}$ 
\footnote{Note that $\beta=2,3$, and 4 correspond to the case $c=0_Z$, $1_Z$, and $0_X$, respectively.}. 
For this, we again use Azuma's inequality~\cite{Azuma1967}, and we obtain 
\begin{align}
  \label{az1}
  &  p\Big(S_{c,1,u,\U{det},\alpha\oplus1,X}-\sum_{i\in \mathcal{S}_{\U{det}}}
    p(n^{i}=1,t^{i}=u,\alpha^{i}=\beta,b^i=X,y^{i}=\alpha\oplus1|{\bm \zeta}^{i-1},y^{i}\neq\emptyset)
    \le -g_{\U{A}}(N_{\U{det}},\epsilon^{c,1,u,\alpha\oplus1,X}_{\U{A}})\Big)\notag\\
&  \le\epsilon^{c,1,u,\alpha\oplus1,X}_{\U{A}}
\end{align}
and
\begin{align}
  \label{az2}
  &  p\Big(S_{c,1,u,\U{det},\alpha\oplus1,X}-\sum_{i\in \mathcal{S}_{\U{det}}}
    p(n^{i}=1,t^{i}=u,\alpha^{i}=\beta,b^i=X,y^{i}=\alpha\oplus1|{\bm \zeta}^{i-1},y^{i}\neq\emptyset)
    \ge g_{\U{A}}(N_{\U{det}},\epsilon^{c,1,u,\alpha\oplus1,X}_{\U{A}})\Big)\notag\\
&  \le\epsilon^{c,1,u,\alpha\oplus1,X}_{\U{A}}.
\end{align}
Now, if we combine Eqs.~(\ref{a2}), (\ref{transdecom}), (\ref{az1}) and (\ref{az2}),
we obtain the upper bound on $N_{\U{ph},Z,Z,1,u,\U{det}}$ in Eq.~(\ref{phaseerror}) as
\begin{align}
  N_{\U{ph},Z,Z,1,u,\U{det}}\le\sum^1_{\alpha=0}p^{\U{U}}_{\alpha}\sum_{c\in\mathcal{C}}\Gamma_{\alpha, c}^{\U{U}}
  \frac{S_{c,1,u,\U{det},\alpha\oplus1,X}+\U{sgn}(\Gamma_{\alpha,c}^{\U{U}}) g_{\U{A}}(N_{\U{det}},\epsilon^{c,1,u,\alpha\oplus1,X}_{\U{A}})}
       {p(c)p^B_X}+g_{\U{A}}(N_{\U{det}},\epsilon^{\U{ph},Z,1,u}_{\U{A}})
    \label{Nph}
\end{align}
except for error probability 
\begin{align}
  \epsilon^1_{\U{PH}}:=\sum^1_{\alpha=0}\sum_{c\in\mathcal{C}}\epsilon^{c,1,u,\alpha,X}_{\U{A}}+\epsilon^{\U{ph},Z,1,u}_{\U{A}}.
\label{failurephaseazuma}
\end{align}
Recall that $S_{c,1,u,\U{det},\alpha\oplus1,X}$ in Eq.~(\ref{Nph}) is the number of instances
where Alice emits an untagged single-photon using the setting $\{c\in\mathcal{C}, b=X\}$ 
and Bob obtains a detection event with the measurement outcome $\alpha\oplus1\in\{0,1\}$.
These quantities are not directly observed in the experiments,
and hence we need to estimate them with the decoy-state method, as we have explained in Appendix~\ref{sec:appC}.
Importantly, this has to be done in such a way that one obtains an upper-bound on $N_{\U{ph},Z,Z,1,u,\U{det}}$. 
In the case of $S_{c,1,u,\U{det},\alpha\oplus1,X}$,
we take the following upper or lower bounds, depending on the sign of $\Gamma^{\U{U}}_{\alpha,c}$,
such that $N_{\U{ph},Z,Z,1,u,\U{det}}$ takes its upper bound:
\begin{align}
  S^{\U{U}}_{c,1,u,\U{det},\alpha\oplus1,X}~~\U{if}~~\Gamma^{\U{U}}_{\alpha,c}>0,
  \label{singlegain1}\\
S^{\U{L}}_{c,1,u,\U{det},\alpha\oplus1,X}~~\U{if}~~\Gamma^{\U{U}}_{\alpha,c}\le0,
\label{singlegain2}
\end{align}
where $S^{\U{U}}_{c,1,u,\U{det},\alpha\oplus1,X}$ and $S^{\U{L}}_{c,1,u,\U{det},\alpha\oplus1,X}$ are defined in Eqs.~(\ref{eq:resultc1U}) and
(\ref{eq:resultc1L}), respectively. By substituting Eqs.~(\ref{singlegain1}) and (\ref{singlegain2}) into the rhs of Eq.~(\ref{Nph}),
we obtain $N^{\U{U}}_{\U{ph},Z,Z,1,u,\U{det}}$. 

Finally, we calculate the failure probability associated to the
estimation of the upper bound on $N_{\U{ph},Z,Z,1,u,\U{det}}$ conditioned on $n_{\U{tag}}\le N_{\U{tag}}$ and $S_{\U{det}}=N_{\U{det}}$,
which we denote by $\epsilon_{\U{PH}}$. 
It is given by the sum of $\epsilon^1_{\U{PH}}$ in Eq.~(\ref{failurephaseazuma}) and $\epsilon^{2}_{\U{PH}}$ that is the sum
of the failure probabilities associated to the estimation of $\{S_{c,1,u,\U{det},\alpha,X}\}_{\alpha=0,1,c\in\mathcal{C}}$:
\begin{align}
  \epsilon_{\U{PH}}=\epsilon^{1}_{\U{PH}}+\epsilon^{2}_{\U{PH}}.
\end{align}
The parameter $\epsilon^{2}_{\U{PH}}$ has the form
$\epsilon^{2}_{\U{PH}}=\sum_{\alpha=0,1}\sum_{c\in\mathcal{C}}\epsilon^{c,1,u,\alpha,X}$ with $\epsilon^{c,1,u,\alpha,X}$ being
the failure probability associated to the estimation of $S_{c,1,u,\U{det},\alpha,X}$ in Eq.~(\ref{singlegain1}) or Eq.~(\ref{singlegain2}),
namely, $\epsilon^{c,1,u,\alpha,X}=\epsilon^{c,1,u,\alpha,X}_{\U{U}}$ or
$\epsilon^{c,1,u,\alpha,X}=\epsilon^{c,1,u,\alpha,X}_{\U{L}}$ depending on whether we use the upper or the lower bound.

\subsection{Derivation of $N_{\U{ph},Z,Z,1,u,\U{det}}$ with general phase intervals in Eq.~(\ref{Rph})}
\label{apEgeneralRph}
In this subsection, we describe the expression of the upper bound on the number of phase errors 
$N_{\U{ph},Z,Z,1,u,\U{det}}$ with the general phase intervals that are shown in Eq.~(\ref{Rph}):
\begin{align}
R^{0_Z}_{\U{\U{ph}}}=[\theta^{\U{L}}_{0_Z},\theta^{\U{U}}_{0_Z}],~~
R^{1_Z}_{\U{\U{ph}}}=[\theta^{\U{L}}_{1_Z},\theta^{\U{U}}_{1_Z}],~~
R^{0_X}_{\U{\U{ph}}}=[\theta^{\U{L}}_{0_X},\theta^{\U{U}}_{0_X}],
\end{align}
where $-\frac{\pi}{6}<\theta^{\U{L}}_{0_Z}\le0$, $0\le\theta^{\U{U}}_{0_Z}<\frac{\pi}{6}$,
$\frac{5\pi}{6}<\theta^{\U{L}}_{1_Z}\le\pi$, $\pi\le\theta^{\U{U}}_{1_Z}<\frac{7\pi}{6}$,
$\frac{\pi}{3}<\theta^{\U{L}}_{0_X}\le\frac{\pi}{2}$, and $\frac{\pi}{2}\le\theta^{\U{U}}_{0_X}<\frac{2\pi}{3}$. 
In the estimation of the number of phase errors, the difference of the phase interval 
is only reflected in the upper bounds 
$\{\Gamma^{\U{U}}_{\alpha,c}\}_{\alpha=0,1,c\in\mathcal{C}}$ on $\{\Gamma^i_{\alpha,c,g^i}\}_{\alpha=0,1,c\in\mathcal{C}}$ 
in Eq.~(\ref{gs}) and $\{p^{\U{U}}_{\alpha}\}_{\alpha=0,1}$ in Eq.~(\ref{palphaU}), and hence 
we just need to replace $\{\Gamma^{\U{U}}_{\alpha,c}\}_{\alpha=0,1,c\in\mathcal{C}}$ and $\{p^{\U{U}}_{\alpha}\}_{\alpha=0,1}$ in Eq.~(\ref{Nph}).
The explicit replacements can be obtained by considering partial difference functions of
$\Gamma^i_{\alpha,c,g^i}$ in Eq.~(\ref{gs}) with respect to $\{\theta^i_{c,g^i}\}_{c\in\mathcal{C}}$ and by deriving 
the parameters $\{\theta^i_{c,g^i}\}_{c\in\mathcal{C}}$ to achieve the maximum of $\Gamma^i_{\alpha,c,g^i}$.
$\{p^{\U{U}}_{\alpha}\}_{\alpha=0,1}$ can be obtained in a similar way. 
The results are summarised as follows.
$$
\Gamma^{\U{U}}_{0,0_Z}\to \Gamma_{0,0_Z,g^i}^{i}(\theta^{\U{L}}_{0_Z},\theta^{\U{L}}_{1_Z},\theta^{\U{U}}_{0_X}),~
  \Gamma^{\U{U}}_{0,1_Z}\to \Gamma_{0,1_Z,g^i}^{i}(\theta^{\U{U}}_{0_Z},\theta^{\U{U}}_{1_Z},\theta^{\U{L}}_{0_X}),~
  \Gamma^{\U{U}}_{0,0_X}\to \max_{x,y,z\in\{\U{L},\U{U}\}}\{\Gamma_{0,0_X,g^i}^{i}(\theta^x_{0_Z},\theta^y_{1_Z},\theta^z_{0_X})\},$$
$$
  \Gamma^{\U{U}}_{1,0_Z}\to \Gamma_{1,0_Z,g^i}^{i}(\theta^{\U{U}}_{0_Z},\theta^{\U{L}}_{1_Z},\theta^{\U{L}}_{0_X}),~
  \Gamma^{\U{U}}_{1,1_Z}\to \Gamma_{1,1_Z,g^i}^{i}(\theta^{\U{U}}_{0_Z},\theta^{\U{L}}_{1_Z},\theta^{\U{U}}_{0_X}),$$
$$
  \Gamma^{\U{U}}_{1,0_X}\to
  \begin{cases}
    \Gamma_{1,0_X,g^i}^{i}(\theta^{\U{L}}_{0_Z},\theta^{\U{U}}_{1_Z},\theta^{\U{U}}_{0_X})&
  \U{if}~~\theta^{\U{U}}_{0_X}<\frac{\theta^{\U{L}}_{0_Z}+\theta^{\U{U}}_{1_Z}}{2},\\
  \Gamma_{1,0_X,g^i}^{i}\left(\theta^{\U{L}}_{0_Z},\theta^{\U{U}}_{1_Z},\frac{\theta^{\U{L}}_{0_Z}+\theta^{\U{U}}_{1_Z}}{2}\right)&
  \U{if}~\frac{\theta^{\U{L}}_{0_Z}+\theta^{\U{U}}_{1_Z}}{2}\in R^{0_X}_{\U{ph}},\\
  \Gamma_{1,0_X,g^i}^{i}(\theta^{\U{L}}_{0_Z},\theta^{\U{U}}_{1_Z},\theta^{\U{L}}_{0_X})&\U{if}~~
  \frac{\theta^{\U{L}}_{0_Z}+\theta^{\U{U}}_{1_Z}}{2}<\theta^{\U{L}}_{0_X},\\
\end{cases}
$$
\begin{align}
  p^{\U{U}}_{0} &\to \frac{1+\cos\frac{\theta^{\U{U}}_{0_Z}-\theta^{\U{L}}_{1_Z}}{2}}{2},~p^{\U{U}}_{1}\to \frac{1-\cos\frac{\theta^{\U{L}}_{0_Z}-\theta^{\U{U}}_{1_Z}}{2}}{2}.
  \end{align}

\end{document}